\documentclass[prb,longbibliography,showpacs,onecolumn,superscriptaddress,amsmath,amssymb,verbatim]{revtex4-2}
\usepackage{graphicx}% Include figure files
\usepackage{dcolumn}% Align table columns on decimal point
\usepackage{bm}% bold math
\usepackage{csquotes}
\usepackage{setspace}
\usepackage[pdftex,colorlinks=true,citecolor=blue,linkcolor=blue,urlcolor=blue,bookmarks=true]{hyperref}

\usepackage{soul}        % for \hl{}
\sethlcolor{yellow} 

\renewcommand{\keywords}[1]{%
  \vspace{1em}
  \noindent\textbf{Keywords:} #1
  \vspace{1em}
}

\newenvironment{methods}{
    \section*{Methods} % unnumbered section
}{}

\begin{document}
	\preprint{APS}

\title{Electrical Control of Excitons  in Bare-MoSe$_2$ and MoSe$_2$/NbSe$_2$ Heterostructure}

\author{Atanu Patra}
\email[]{atanu.patra@uni-wuerzburg.de}
\affiliation{Julius-Maximilians-Universität Würzburg, Physikalisches Institut, Lehrstuhl für Technische Physik, Am Hubland, 97074 Würzburg, Germany}

\author{Vishakha Kaushik}
%\email[]{}
\affiliation{Julius-Maximilians-Universität Würzburg, Physikalisches Institut, Lehrstuhl für Technische Physik, Am Hubland, 97074 Würzburg, Germany}

\author{Ali Sepas}
\affiliation{Department of Materials and Production, Aalborg University, DK-9220 Aalborg Øst, Denmark}

\author{Subhamoy Sahoo}
\affiliation{Julius-Maximilians-Universität Würzburg, Physikalisches Institut, Lehrstuhl für Technische Physik, Am Hubland, 97074 Würzburg, Germany}

\author{Mathias Federolf}
\affiliation{Julius-Maximilians-Universität Würzburg, Physikalisches Institut, Lehrstuhl für Technische Physik, Am Hubland, 97074 Würzburg, Germany}

\author{Christian G. Mayer}
\affiliation{Julius-Maximilians-Universität Würzburg, Physikalisches Institut, Lehrstuhl für Technische Physik, Am Hubland, 97074 Würzburg, Germany}
\affiliation{Physikalisches Institut and Würzburg-Dresden Cluster of Excellence ct.qmat, Germany}

\author{Sebastian Klembt}
\affiliation{Julius-Maximilians-Universität Würzburg, Physikalisches Institut, Lehrstuhl für Technische Physik, Am Hubland, 97074 Würzburg, Germany}
\affiliation{Physikalisches Institut and Würzburg-Dresden Cluster of Excellence ct.qmat, Germany}

\author{Monika Emmerling}
\affiliation{Julius-Maximilians-Universität Würzburg, Physikalisches Institut, Lehrstuhl für Technische Physik, Am Hubland, 97074 Würzburg, Germany}

\author{Simon Betzold}
\affiliation{Julius-Maximilians-Universität Würzburg, Physikalisches Institut, Lehrstuhl für Technische Physik, Am Hubland, 97074 Würzburg, Germany}

\author{Seth Ariel Tongay}
\affiliation{Materials Science and Engineering, School for Engineering of Matter, Transport and Energy, Arizona State University, Tempe, 85287 Arizona, United States}

\author{Fabian Hartmann}
\affiliation{Julius-Maximilians-Universität Würzburg, Physikalisches Institut, Lehrstuhl für Technische Physik, Am Hubland, 97074 Würzburg, Germany}

\author{Thomas Garm Pedersen}
\email[]{tgp@mp.aau.dk}
\affiliation{Department of Materials and Production, Aalborg University, DK-9220 Aalborg Øst, Denmark}

\author{Sven Höfling}
\email[]{sven.hoefling@uni-wuerzburg.de}
\affiliation{Julius-Maximilians-Universität Würzburg, Physikalisches Institut, Lehrstuhl für Technische Physik, Am Hubland, 97074 Würzburg, Germany}
\affiliation{Physikalisches Institut and Würzburg-Dresden Cluster of Excellence ct.qmat, Germany}

%\date{\today}
%\keywords{hBN,  $V_B^-$-spin-defects, Raman, photoluminescence, polarization-dependent}

\begin{abstract}
Monolayer transition metal dichalcogenides (TMDCs) are promising materials for next-generation optoelectronic devices, owing to their strong excitonic responses and atomic thickness. Controlling their light emission electrically is a crucial step towards realizing practical nanoscale optoelectronic devices such as light-emitting diodes and optical modulators. 
%Electrical control of their light emission is therefore a key step toward practical applications  such as nanoscale light-emitting diodes and optical modulators. 
However, photoluminescence (PL) quenching in van der Waals TMDC/metal heterostructures, caused by ultrafast interlayer charge or energy transfer, impedes such electrical modulation. %However, achieving this control is challenging because van der Waals heterostructures with metals often suffer from photoluminescence (PL) quenching due to ultrafast interlayer charge or energy transfer.
Here, we investigate monolayer-MoSe$_2$/bulk-NbSe$_2$ heterostructures and demonstrate that a vertical electric field can effectively recover the PL intensity up to $\sim$ 80$\%$ of bare-MoSe$_2$. Furthermore, our analysis reveals that the room temperature PL intensity can be tuned by nearly three orders of magnitude in bare-MoSe$_2$ and by about one order of magnitude in MoSe$_2$/NbSe$_2$ heterostructures. First-principles calculations incorporating spin-orbit coupling reveal that the perpendicular electric fields drive a transition from a direct to an indirect bandgap, fundamentally altering the optical response in the heterostructure. Unlike bare-MoSe$_2$, the heterostructure exhibits a pronounced thermal dependence of the enhancement factor, implying that exciton lifetime dominates over interfacial transfer processes. Our findings demonstrate reversible, electric-field-driven PL control at a TMDC/metal interface, providing a pathway to electrically tunable light emission and improved contact engineering in two-dimensional optoelectronic devices.
\end{abstract}
\maketitle
\section{Introduction}
%\textbf{Introduction}

Van der Waals (vdW) heterostructures of transition metal dichalcogenides (TMDCs) materials have enabled the creation of artificial systems with novel functionalities that do not exist in nature. These include Moiré superlattices arising from controlled twisting or lattice mismatch, allowing for the exploration of emergent, strongly correlated electronic phases \cite{tong2017topological,li2021continuous,xiong2023correlated,ma2021strongly,xu2024hydrodynamic}. Furthermore, vdW stacking enables tunable type-I and type-II band alignment through the selection of material combination, stacking angle, and interlayer coupling \cite{meng2020electrical,ubrig2020design}. They also facilitate the realization of long-lived interlayer excitons \cite{rivera2015observation} with spatially separate charge carriers, distinct from the ultrafast recombination seen in monolayers \cite{robert2016exciton,palummo2015exciton}. Such features are promising for advanced optoelectronic and excitonic device platforms~\cite{liu2019room,gonzalez2022room,meng2023photonic}.
However, the integration of TMDCs into devices remains challenging due to the high contact resistance at the TMDC/metal interface. Although graphene is widely used as a transparent electrode, its strong interlayer coupling with TMDCs induces significant interfacial charge and energy transfer, resulting in severe photoluminescence (PL) quenching \cite{yang2018effect,lorchat2020filtering}. Moreover, work function mismatch between metal contacts and semiconducting TMDCs further contributes to high contact resistance \cite{liu2016van}, thereby constraining device performance. These limitations necessitate the exploration of alternative layered metals that offer better electronic compatibility. A highly interesting candidate is the layered two-dimensional (2D) metal NbSe$_2$, which is a type-II superconductor at temperatures below 8 K and metallic at higher temperatures with coexisting charge density wave order around 32 K \cite{xi2015strongly}. Recent studies have shown the integration of 2D semiconducting layers into heterostructures with 2D NbSe$_2$ providing a platform for the manipulation of electronic and excitonic properties \cite{joshi2020localized}, and for the development of device applications, such as self-powered photodetectors \cite{li2023self}. In these devices, the dynamic tuning of optical properties is essential, enabling real-time control over excitonic behavior and light–matter interactions, which is critical for developing next-generation programmable and tunable optoelectronic functionalities using 2D materials \cite{mak2016photonics}.  %Importantly, such heterostructures benefit from the ability to electrically tune their excitonic and optical properties, allowing real-time modulation of light–matter interactions, which is critical for next-generation reconfigurable optoelectronic devices \cite{mak2016photonics}.

%In recent years, electrostatic gating offers a non-invasive approach to precisely tune carrier densities in TMDCs \cite{meng2020electrical}. 
In recent years, electrostatic gating offers a non-invasive approach to precisely tune optical properties in TMDCs \cite{meng2020electrical}. This control enables modulation of quasiparticle bandgaps \cite{chu2015electrically}, phase transitions \cite{ramasubramaniam2011tunable,kang2017universal}, and symmetry-breaking phenomena \cite{jones2013optical,wu2013electrical}, as well as excitonic states \cite{jauregui2019electrical, kistner2024electric,li2015electric}, all without introducing structural disorder, unlike chemical doping.
Vertical electric fields also alter hybrid excitons with strong dipolar interactions in TMDC bilayers \cite{leisgang2020giant,lorchat2021excitons,shi2022bilayer}. Moreover, a recent study predicts that electric fields can modulate band alignment and Schottky barrier heights at TMDC/metal interfaces \cite{lv2018tunable}, a key factor in minimizing contact resistance and enhancing device performance. Building on these studies, understanding how vertical electric fields govern excitonic behavior and interfacial charge dynamics in TMDC/2D-metal heterostructures, especially under varying thermal conditions, remains largely \mbox{unexplored}.
%Despite these advances, a comprehensive understanding of how vertical electric fields influence excitonic behavior and interfacial charge dynamics, particularly in TMDC/2D-metal heterostructures under varying thermal conditions is still lacking.

In this work, we leverage vertical electric field modulation to investigate the PL properties in monolayer (ML)-MoSe$_2$/bulk-NbSe$_2$ (TMDC/metal) heterostructure. Measurements were performed at two  positions, namely, bare-MoSe$_2$ (`off' position) and the MoSe$_2$/NbSe$_2$ heterostructure (`on' position). At room temperature ($T =$ 295 K), the PL emission at the `on' position is reduced by more than an order of magnitude compared to the `off' position. %A decrease in PL emission by more than an order of magnitude is observed at room-temperature ($T =$ 295 K) between `on' and `off' position. 
Our results reveal that the direction of the electric field determines the PL enhancement in bare-MoSe$_2$ and MoSe$_2$/NbSe$_2$  heterostructure  with a different degree of enhancement. Notably, at the `on' position, the PL enhancement factor decreases with lowering temperature, while it remains nearly constant at the `off' position. All measurements in this work were carried out down to $T = 20$ K, with a specific focus on the normal (non-superconducting) metallic regime of NbSe$_2$. %In case of the `on' position,  a decrease of PL enhancement factor with reducing temperatures is observed. In contrast, for the `off' position, these values remain unaltered. 
Analytical calculations estimate that the electric field modifies the excitonic oscillator strength by altering the electron-hole (\textit{e-h}) wavefunction overlap for the `off' position. However, first-principle calculations on the MoSe$_2$/NbSe$_2$ heterostructure, representing the `on' position, demonstrate that the field modifies the nature of the bandgap. %, thereby, affecting the optical emission.
Consequently, controlling the electric field allows modulation of the PL intensity and can recover emission at the `on’ position by up to $\sim$ 80$\%$ of the `off’ position. %Controlling the electric field thus modulates PL intensity and recovers emission at the `on’ position up to $\sim$ 80$\%$ comapred to `off’ position.  %These results underscore the interplay between metal-semiconductor work function difference and highlight a competition between  exciton dissociation and exciton recombination. 
%\newpage
\begin{figure}[h]
	\centering
	\includegraphics[width=0.98\textwidth]{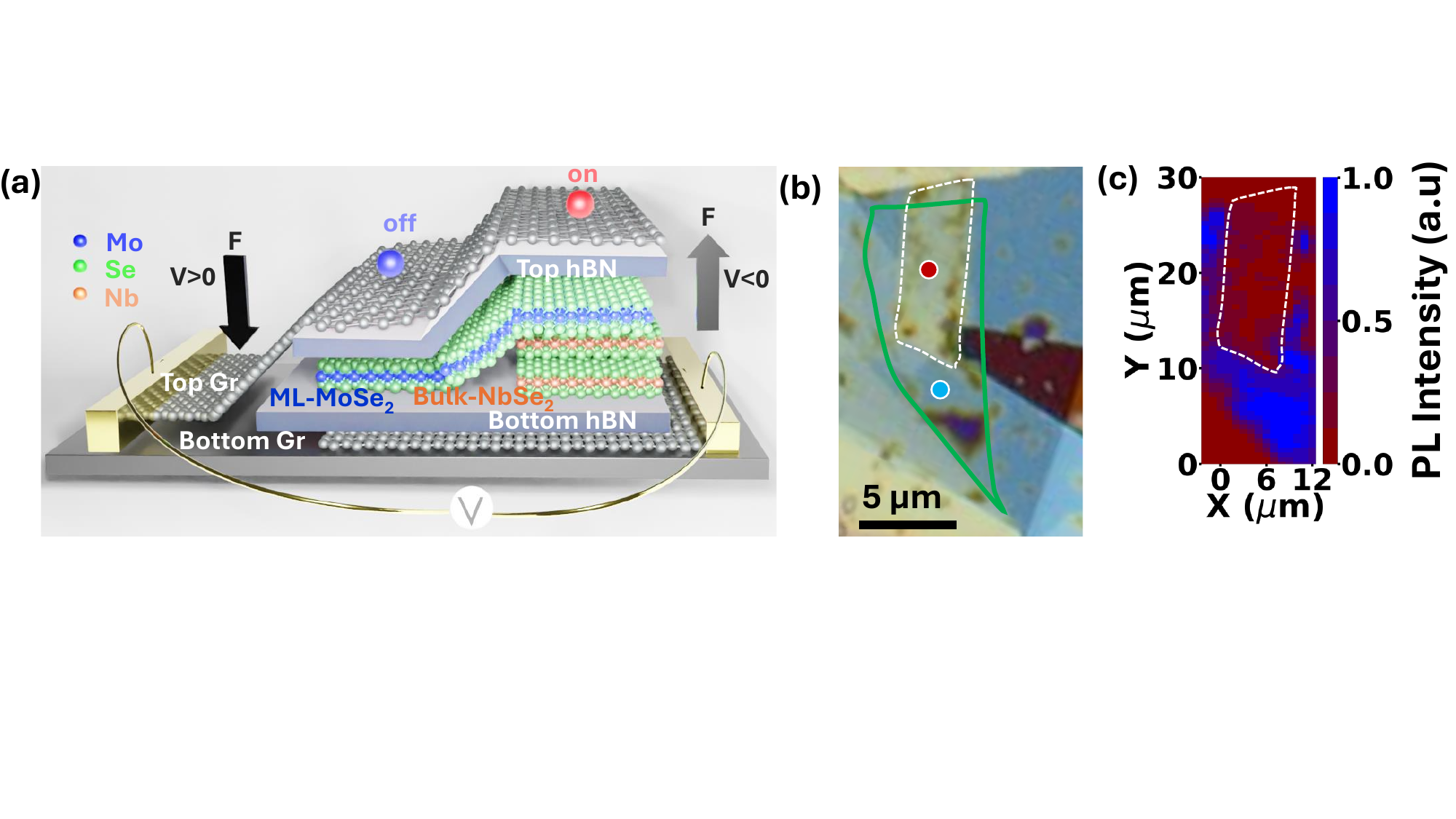}
	\vspace*{-3.0cm}
	\caption{(a) Schematic diagram and (b) optical image of a typical ML-MoSe$_2$/bulk-NbSe$_2$ heterostructure with top and bottom  few-layer graphene contacts. (c) PL  intensity map resulting from ML-MoSe$_2$ at room temperature ($T =$ 295 K).}\label{sample}
\end{figure}

\newpage
\section{Results}\label{sec2}
\noindent\textbf{PL quenching in MoSe$_2$/NbSe$_2$ heterostructure}\\
A schematic and optical micrograph of a  typical MoSe$_2$/NbSe$_2$  stacked heterostructure, comprising ML-MoSe$_2$ interfaced with bulk NbSe$_2$, are shown in Fig. \ref{sample}a and b, respectively. The thicknesses of  MoSe$_2$ and  NbSe$_2$ layers are determined by the optical contrast and Raman measurement, as shown in Supplementary Fig. \ref{fig:Raman}. The heterostructure, marked by a white contour in Fig. \ref{sample}b, is formed in the overlapping region of the two constituent materials. Note that the layers were not rotationally aligned, implying weak electronic hybridization between the incommensurate semiconducting and metallic layers. To study the effect of electric fields in the heterostructure, a capacitive structure is assembled on a pre-patterned gold electrode (details given in Supplementary section) using a layer-by-layer dry transfer stacking technique with hexagonal boron nitride (hBN) as the dielectric layer. The stack consists of few-layer-graphene/hBN/NbSe$_2$/MoSe$_2$/hBN/few-layer-graphene, with both the top and bottom hBN layers measuring approximately 12 $\pm$ 3 nm in thickness. 
The spatial PL intensity map, as shown in Fig. \ref{sample}c, predominantly displays two colors. The blue colored area corresponds to bare-MoSe$_2$,  while the region outlined by the white contour, representing  the MoSe$_2$/NbSe$_2$ heterostructure, appears similar in color to the background. This indicates substantial PL quenching of MoSe$_2$ across the entire heterostructure area involving NbSe$_2$.

\begin{figure}[t]
	\centering
	\includegraphics[width=0.98\textwidth]{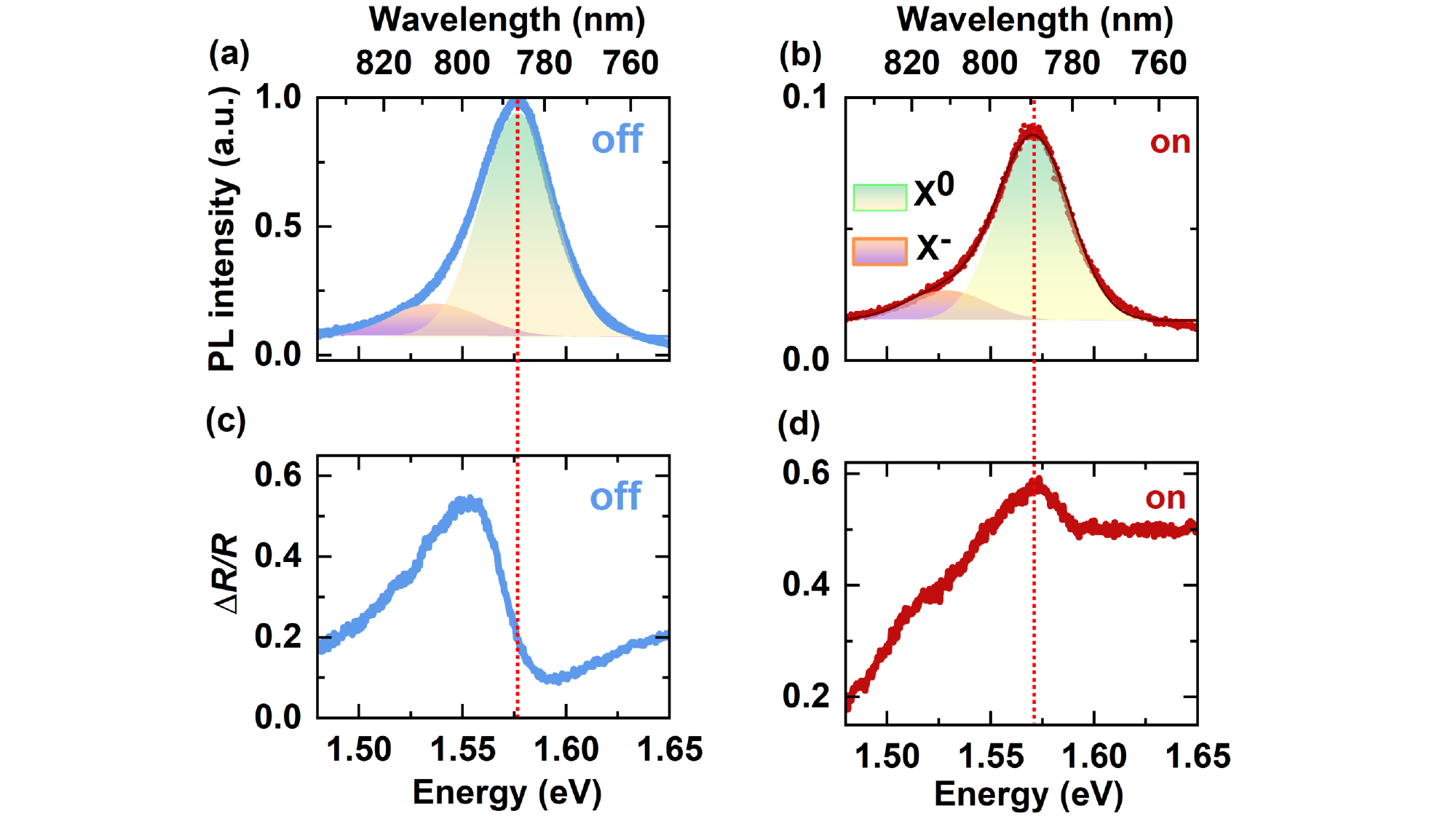}
	\vspace*{-0.0cm}
	\caption{(a) and (b)  PL spectra at $T =$ 295 K measured at the `off' and `on' positions, respectively, revealing a one order of magnitude reduction in intensity at the `on' position compared to the `off' position. (c)  and (d) Corresponding reflectance contrast ($\Delta R/R$) spectra measured at the same points and at the same temperature.}\label{charact.}
\end{figure}

For simplicity, further analysis focuses on two representative positions, denoted as  `off' and `on' in   Fig. \ref{sample}a  by blue and red  points, respectively. At $T =$ 295 K, PL spectra in the absence of a vertical electric field (i.e., voltage, $V$ = 0 V) at these two positions are presented in Fig. \ref{charact.}a and b, respectively. The spectra were fitted with two Gaussian functions representing the A-exciton and trion  at  1.57 $\pm$ 0.05 eV and 1.53 $\pm$ 0.05 eV, respectively.
%The spectra are fitted with two Gaussian functions related to A-excitons and trions. Both spectra exhibit ground-state A-excitons at  1.57$\pm$0.05 eV.
The reflectance contrast spectra, $\Delta R/R=\frac{R-R_0}{R_0}$, where $R$ and $R_0$ are the reflectance signal of the stack, with and without MoSe$_2$, respectively, for both `off' and `on' positions, are shown in  Fig. \ref{charact.}c and d, respectively. The `off'-position spectrum predominantly indicates the real component of the dielectric function, while the `on'-position is governed by both real and imaginary parts \cite{li2014measurement}. The transition energies for the respective positions are marked by dotted lines in the figure.  
Although excitons in ML-MoSe$_2$ are spatially confined within the layer, the associated dipolar electric fields naturally extend beyond the layer due to the reduced dimensionality. As a result, forming a heterostructure of MoSe$_2$  with metallic NbSe$_2$ inherently influences the excitonic behavior by enhancing environmental screening and modifying the local dielectric landscape. This interaction significantly alters the optical response, manifesting as distinct variations in $\Delta R/R$ at these two positions.

The most striking feature is the PL intensity suppression of around one order of magnitude ($\sim$ 11) between `off' and `on' positions. We conducted the measurements using a 532 nm laser at $\sim$ 15 $\mu$W  power to ensure operation within  the linear excitonic regime,  as illustrated in Fig. \ref{fig:power dependence}a. Notably, the quenching factor remains consistent across the experimental power range as depicted in  Fig. \ref{fig:power dependence}b. This quenching phenomenon is consistent with observations at semiconductor/metal interfaces, such as TMDC/graphene heterostructures, where PL intensity is suppressed due to photoinduced charge or energy transfer via Dexter or Förster processes \cite{froehlicher2018charge}. This indicates substantial PL quenching of MoSe$_2$ across the entire heterostructure area involving NbSe$_2$. These processes occur on sub-picosecond timescales, significantly faster than the nanosecond-scale exciton lifetime at $T =$ 295 K. The extent of PL quenching is strongly influenced by interlayer coupling, including factors such as interlayer distance, stacking sequence, and dielectric screening \cite{yang2018effect}.  Since this quenching reflects a reduction in the exciton density, external electric fields that modulate exciton formation and recombination dynamics provide a promising route to restore or enhance PL emission in vdW heterostructures.

%However, the most striking feature is the significant variation in the PL intensity of the A-exciton between the two positions. The spatial PL intensity map, as shown in Fig. \ref{charact.}c, predominantly displays two colors: the brighter areas corresponding to bare MoSe$_2$ and the region outlined by the white contour appearing similar in color to the background. This indicates substantial PL quenching across the entire heterostructure area involving NbSe$_2$. Specifically, measurements at two representative points, marked in blue (bare MoSe$_2$) and red (heterostructure), reveal a PL intensity suppression of around one order of magnitude ($\sim$ 11). This quenching phenomenon is consistent with observations at metal-semiconductor interfaces, such as graphene/TMDC heterostructures, where PL is suppressed due to photoinduced charge or energy transfer via Dexter or Förster processes \cite{froehlicher2018charge}. These processes occur on sub-picosecond timescales, significantly faster than the nanosecond-scale exciton lifetime at room-temperature. The extent of PL quenching is strongly influenced by interlayer coupling, including factors such as interlayer distance, stacking sequence, and dielectric screening \cite{yang2018effect}. Since this quenching reflects exciton dissociation, controlling exciton formation and preventing their dissociation into free carriers via an external electric field could offer a pathway to restore quenched PL in vdW heterostructures.

\begin{figure}[t]
	%\vspace*{-5.0cm}
	\centering
	%\hspace*{-0.5cm}
	\includegraphics[width=1.00\textwidth]{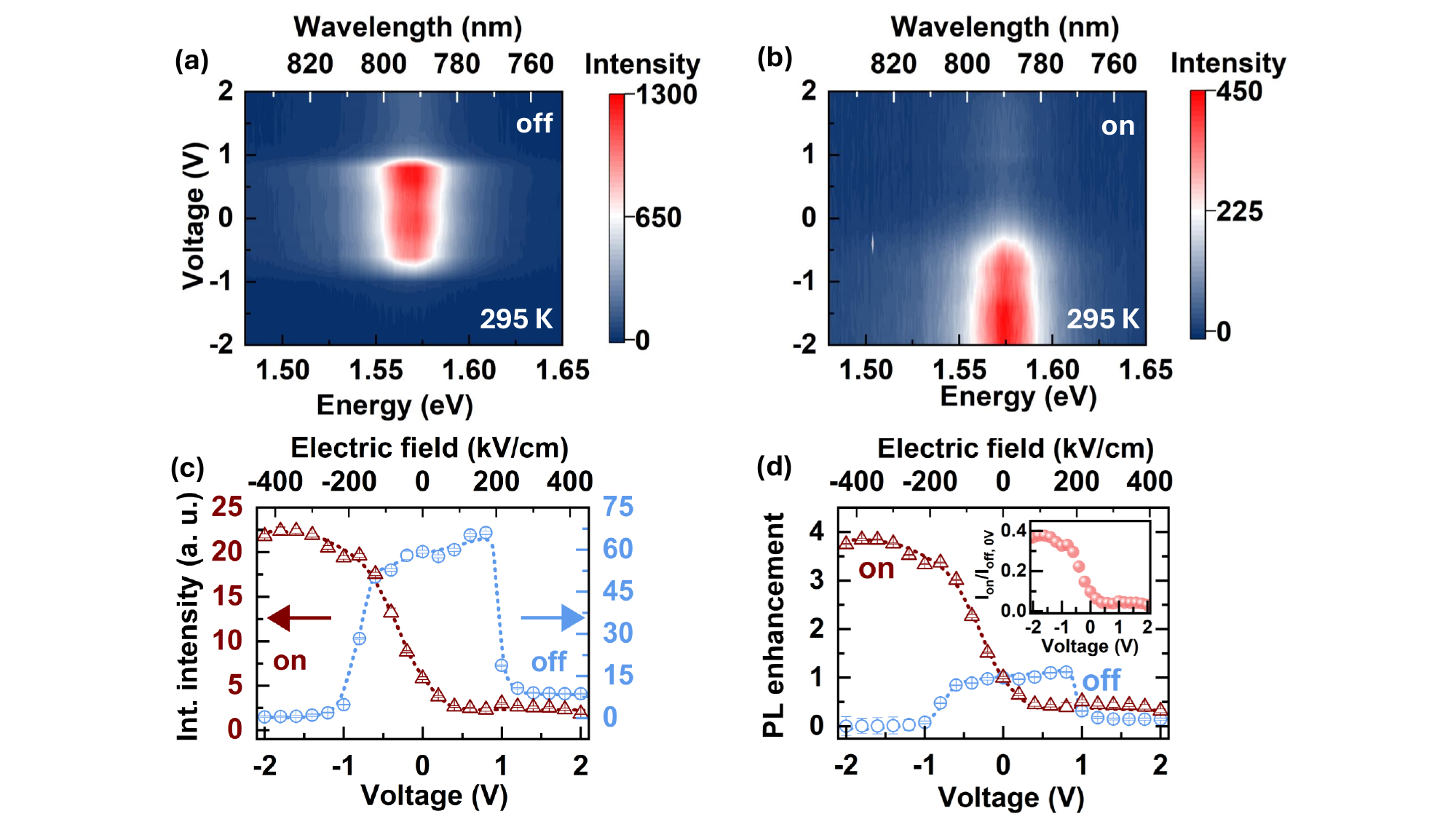}
	%\vspace*{-1.0 cm}
    \caption{(a) and (b) Contour plots of PL spectra as a function of voltage at the `off' and `on' positions, respectively. (c) Extracted energy integrated PL intensity, as described in Fig. \ref{charact.}a and b, with blue circles representing the `off' position and red triangles representing the  `on' position. (d) Enhancement of the integrated intensity relative to its value at $V = 0$ V.  
    Inset: revival of the intensity at the `on' position compared to the `off' position at $V = 0$ V.
    }\label{gate-mapping}
\end{figure}

\noindent\textbf{Electric field-dependent PL enhancement at room temperature}\\
We next investigated the influence of an applied electric field (\textit{F}) on the excitonic behavior, particularly, its effect on spectral intensity, as a proof-of-concept for implementing field-driven tunability in 2D materials-based heterostructures. A positive voltage ($V>0$) corresponds to the top electrode near the MoSe$_2$ being positively biased while the bottom electrode remains grounded, resulting in a downward-pointing electric field, as illustrated in Fig. \ref{sample}a. In contrast, applying a negative voltage ($V<0$) reverses the field direction. Contour color plots of the PL spectra for the `off' and `on' positions are presented in Fig. \ref{gate-mapping}a and b, respectively. 
The integrated intensities at both positions are displayed in Fig. \ref{gate-mapping}c, with blue circles and red triangles representing the `off' and `on' positions, respectively. The integrated intensity exhibits a step-like variation:  the `off' position shows a two-step change, while the `on' position demonstrates a single-step behavior. For the `off' position, the PL intensity increases slightly at $V$ $\sim$ +0.8 V, reaching a maximum before rapidly decreasing at higher voltages. Similarly,  a sharp decrease is observed below $V$ $\sim$ -0.6 V. For the `on' position, a notable enhancement in PL intensity occurs at $V$ $\sim$ -1.8 V. The PL enhancement factor, defined as $\frac{I_V}{I_0}$, where $I_V$ is the PL intensity integrated over energy at a given voltage and $I_0$ is the integrated intensity at zero voltage, is illustrated in Fig. \ref{gate-mapping}d. A modest increase of a factor of $\sim$ 1.1$\times$ is observed for the `off' position, while the `on' position shows a maximum of $\sim$ 4.0$\times$ enhancement. However, at a fixed position, the absolute change in intensity over the applied voltage sweep is much more pronounced, with nearly three orders of magnitude for the `off' position and one order of magnitude for the `on' position, respectively.

%The structural differences previously studied, specifically the much thinner hBN layer used in \cite{wang2023exciton} comparison to that used in the present study, rules out electroluminescence via resonant tunneling of energetic electrons from graphene through hBN to MoSe$_2$ as the source of the enhancement. 
In a previous study, a thin hBN layer was used to achieve resonant tunneling of energetic electrons from graphene to MoSe$_2$ \cite{wang2023exciton}. However, the structural differences in the present work, particularly the use of a thicker hBN layer, rule out this mechanism as the source of the observed PL enhancement. Moreover, the enhancement achieved at different polarities differs between the two positions: $V>0$  for the bare-MoSe$_2$ and $V<0$ for the MoSe$_2$/NbSe$_2$ heterostructure. A direct comparison of the PL intensity between the `on' and `off' positions at $V$=0, is shown as an intensity ratio in the inset of Fig. \ref{gate-mapping}d. This reveals that the electrostatic field revives the optical intensity of the heterostructure to $\sim$40$\%$ of the bare-MoSe$_2$ PL intensity, as shown in the inset of Fig. \ref{gate-mapping}d. The maximum enhancement factor, although, is  different for another similarly structured device (D2) studied, achieving up to 12× enhancement, $\sim$ 80$\%$ of the bare-MoSe$_2$ intensity, as illustrated in  Fig. \ref{fig:V$_g$_Device_NbSe$_2$}. Interlayer coupling between MoSe$_2$ and NbSe$_2$, as well as the overall sample quality, plays a crucial role in this behavior.

Notably, different polarities of $V$ are required to achieve enhancement at the two positions. To further substantiate this finding, we replaced NbSe$_2$ with a few-layer graphene atop MoSe$_2$ in third device D3. In this configuration, PL enhancement was observed for both bare-MoSe$_2$ and graphene/MoSe$_2$ heterostructure at $V>0$, as shown in  Fig. \ref{fig:V$_g$_Gr/MoSe$_2$}. The maximum enhancement factors are $\sim 1.2\times$ for bare-MoSe$_2$ and $\sim 3.8\times$ for graphene/MoSe$_2$ heterostructure, respectively. The result suggests a PL recovery of $\sim$ 23 $\%$ for `on-graphene' heterostructure position compared to the bare-MoSe$_2$. %This behavior is consistent with reports that the interlayer distance in Gr/TMDCs/hBN systems is larger than in WSe\(_2\)/hBN/Gr systems, due to stronger van der Waals interactions between Gr and hBN\cite{yang2018effect}, which increase the interlayer separation. 
Therefore, regardless of the metal used, the observed enhancement and its polarity-dependent behavior are governed by the metal’s position relative to ML-TMDCs and the direction of the applied voltage, respectively.\\
%\newpage
\noindent\textbf{Temperature dependence}  \\
The exciton lifetime of bare-MoSe$_2$ is a few-nanoseconds at $T =$ 295 K and a few-picoseconds  at low temperatures ($T =$ 20 K), as presented in Fig. \ref{fig:Time-resolved-PL}a and b, respectively, and also reported earlier \cite{robert2016exciton,fang2019control}. This results in significant changes in exciton luminescence with varying temperature. 
\begin{figure}[t]
	\centering
	\hspace*{-3.5 cm}
    \vspace*{-0.0 cm}
	\includegraphics[width=1.4\textwidth]{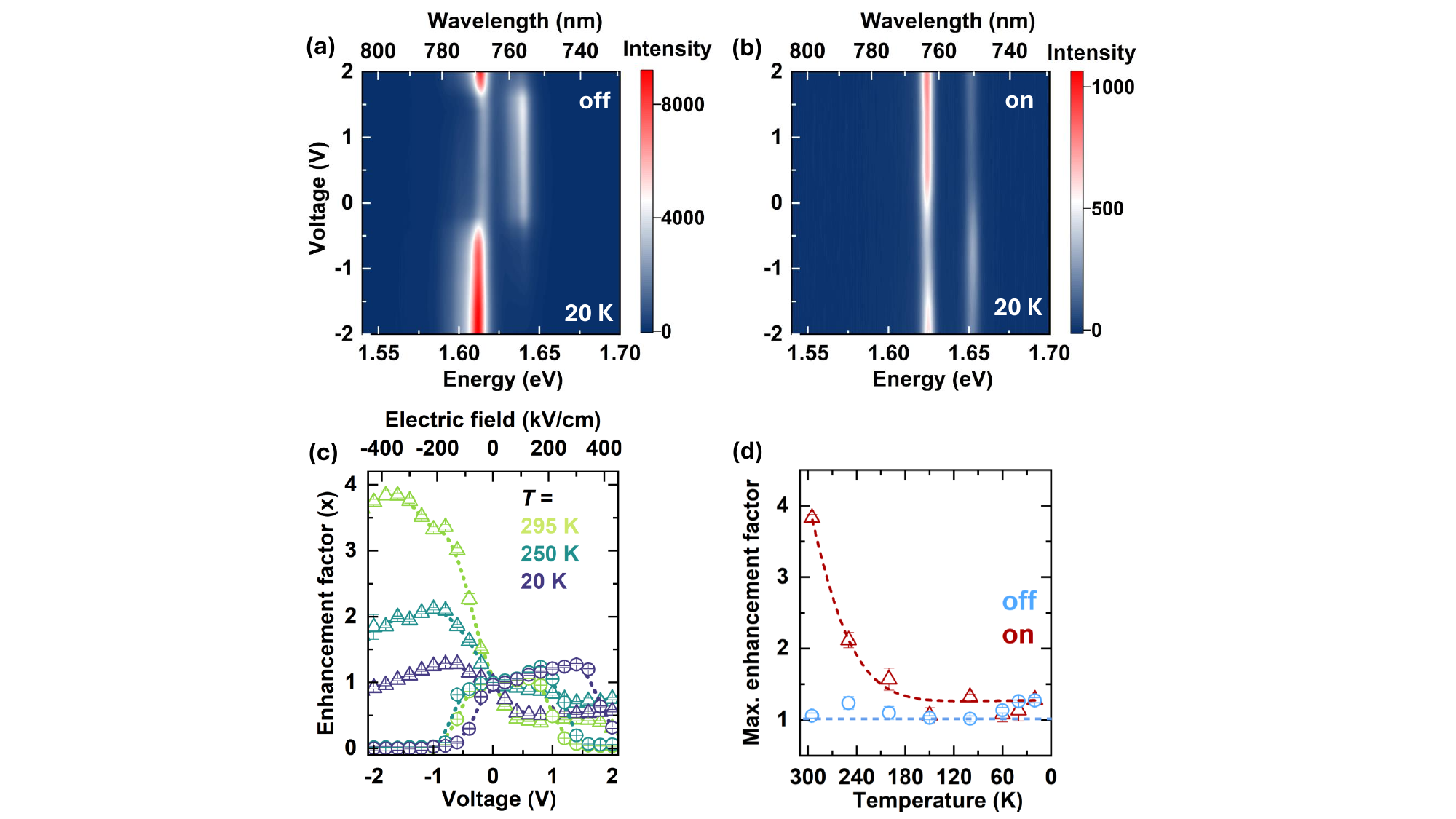}
	\caption{Variation of PL emission with  voltage at $T =$ 20 K for (a) `off' and (b) `on' position. (c) Enhancement for `off' position (circle) and `on' position (triangle) at selective temperatures. (d) Observed maximum enhancement at each temperature at these points. Dashed lines are a guide to the eye.}\label{LT-PL}
\end{figure}
To investigate the impact of radiative lifetime of exciton on PL quenching and their modification under an applied electric field in MoSe$_2$/NbSe$_2$ heterostructure under different thermal conditions, we conducted voltage-dependent studies at various temperatures.
%To investigate the interplay between  PL quenching, and their modification under an applied electric field in MoSe$_2$/NbSe$_2$ heterostructure, we conducted voltage-dependence studies at various temperatures. 
For comparison, contour plots for both the positions at $T =$ 20 K are presented in Fig. \ref{LT-PL}a and b.  Both the plots reveal two prominent peaks corresponding to excitons at $\sim$1.65 eV and trions at $\sim$1.62 eV. Although the mass action law predicts a constant trion-to-exciton intensity ratio in the absence of intentional doping, we observe a strong voltage dependence, likely due to photo-induced doping effects \cite{verzhbitskiy2019suppressed, zhang2018impact, roch2018quantum,lundt2018interplay}. To maintain consistency with the present study's theme, we focus on the exciton behavior.
In contrast to $T =$ 295 K, the PL-enhancement factor of the exciton for the `on' position reduces at $T =$ 20 K, while maintaining a qualitatively similar voltage-dependence. The maximum enhancements at specific temperatures for both positions are shown in Fig. \ref{LT-PL}c, with circles and triangles representing `off' and `on' positions, respectively. %Data positions of open circle represent the `on'  position, and data positions with a filled circle represent the `off' position. 
 Fig. \ref{LT-PL}d demonstrates that the maximum enhancement factor depends on temperature. %Intriguingly, besides enhancement factor, the voltage corresponding to the maximum enhancement ($V^m$) also varies with temperature, as shown in Fig. \ref{LT-PL}e. 
 Notably, the maximum enhancement factor at the `off' position remains $(\sim 1.1 \pm 0.1)$, independent of temperature. In contrast, the maximum enhancement factor for the `on' position decreases continuously from $\sim$ 4.0 to 1.2 as the temperature is lowered from 295 K to 20 K.\\
%In NbSe$_2$, free-electron-like behavior is reported below 1 eV \cite{bachmann1971optical}, while two interband transitions are observed at 1.95 eV and 2.33 eV (at room-temperature), possibly associated with spin-orbit splitting \cite{consadori1971anisotropy}. Based on the typical temperature dependence observed in MoSe$_2$, where the optical gap shifts by about 0.08 eV, a corresponding variation of approximately 0.1 eV in the interband transition energy of NbSe$_2$ is expected. However, the experimentally observed temperature-dependent shift in $V^m$ is around 1.6 V. Furthermore, consistent with expectations for van der Waals heterostructures, theoretical calculations\cite{joshi2020localized,lv2018tunable} suggest no hybridization at the K point in MoSe$_2$/NbSe$_2$ heterostructures. Therefore, temperature-induced variations in the interband gap can be ruled out as the underlying cause of the $V^m$–T dependence.
%The electric field has negligible influence on the electronic phase transition of NbSe$_2$, particularly at room-temperature \cite{xi2016gate}, the observed enhancement factors for the heterostructure and bare MoSe$_2$, hence, cannot originate from changes in NbSe$_2$ alone. 
%Bohr radius ($\sim$ 0.7 nm)\cite{wang2015exciton}
%\noindent\textbf{Discussion}\\
\section{Discussion}
\begin{figure}[t]
	%\vspace*{-5.0cm}
	\centering
	%\hspace*{-0.5cm}
	\includegraphics[width=1.00\textwidth]{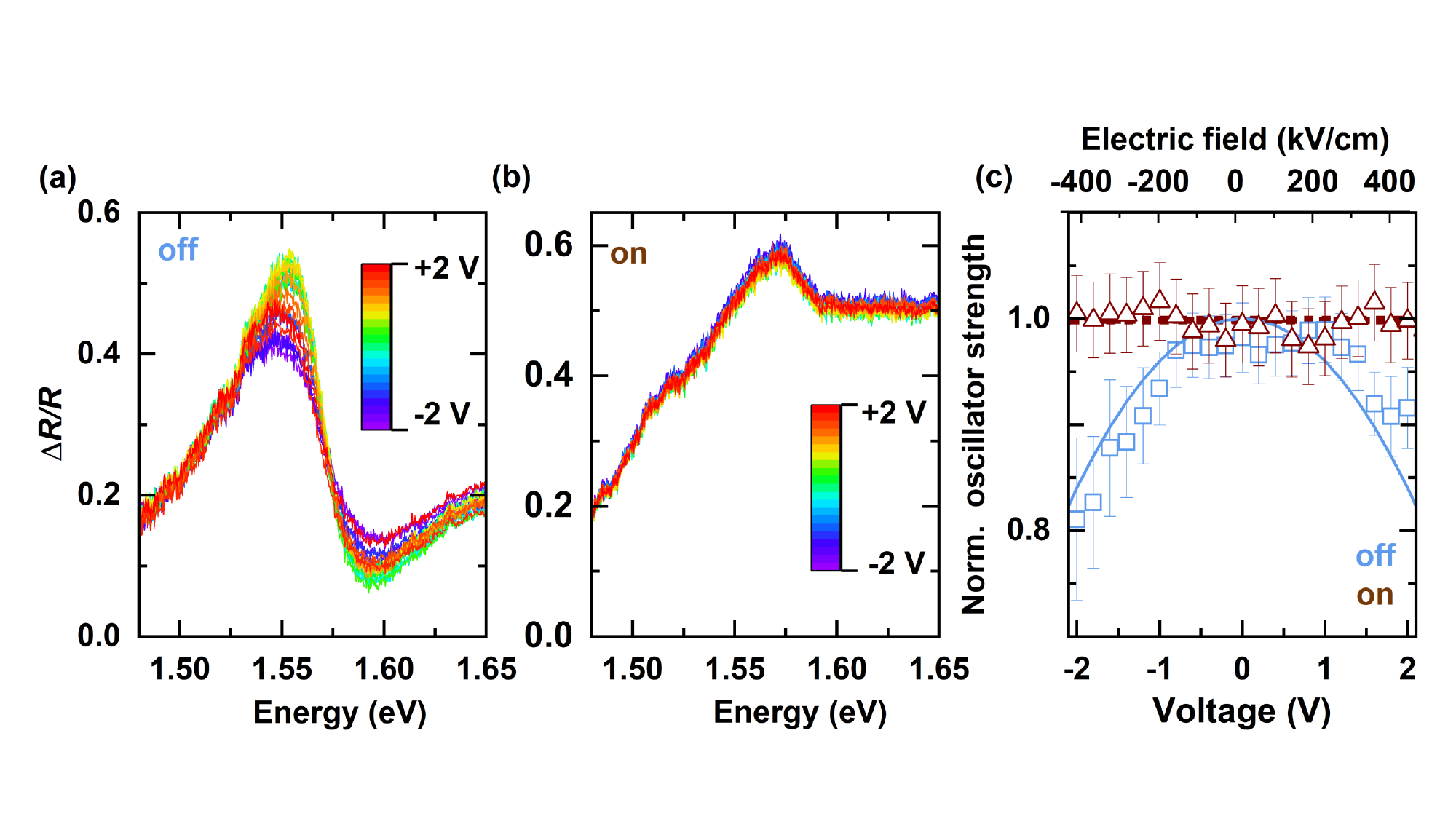}
	\vspace*{-1.0 cm}
    \caption{(a) and (b) Reflectance contrast ($\Delta R/R$) plots at the `off' and  `on'  position with voltage at $T =$ 295 K. (c) Variation in normalized oscillator strength at these positions.}\label{DR_field}
\end{figure}

To explain the field dependence of PL at the `off'  and `on' positions we carried out a detailed analysis of $\Delta R/R$ spectra at these two points as shown in Fig. \ref{DR_field}a and b, respectively, at $T =$ 295 K. A clear vertical electric field dependence  in  $\Delta R/R$ spectra at `off' position can be seen. 
%The $\Delta R/R$ spectra at `off' and `on' positions (see Fig. \ref{charact.}c and d) clearly indicate a strong environmental effect on the A exciton of MoSe$_2$.  To further clarify the role of the vertical electric field in the MoSe$_2$/NbSe$_2$ \hl{heterostructure, and to separate its effect from that of the dielectric environment, we plot the voltage-dependent} $\Delta R/R$ spectra at the 'off' and 'on' positions in Fig. \ref{gate-mapping}e and f, respectively. %To further elucidate the impact of electric field on the dielectric  environment and relative contribution to the optical properties, we have plotted voltage-dependent $\Delta R/R$ for `off' and `on' positions in Fig. \ref{gate-mapping}e and f respectively. 
For a more quantitative understanding, we have fitted each experimental spectrum using the Faddeeva function \cite{raja2019dielectric} (see  Fig. \ref{fig:DR_fit}) to extract the exciton peak position and oscillator strength ($f$).
%To extract the exciton peak position and oscillator strength ($f$), a fit of each experimental spectrum is performed using the Faddeeva function \cite{raja2019dielectric} (shown in Supplementary Fig. \ref{fig:DR_fit}). 
For ML TMDC, due to the lack of inversion symmetry and the absence of a net exciton dipole moment ($\Delta\mu=0$), the field dependence of excitons exhibits  a quadratic behavior, as shown in  Fig. \ref{fig:RT_Stark shift}. This yields an exciton polarizability ($\alpha$) of $(1.3\pm 0.07)\times$ $10^{-18}$ eV$\cdot$ (m/V)$^2$. %In this case, the exciton peak position and oscillator strength are obtained from the fitting, as illustrated earlier in Fig. \ref{charact.}e and g. %This yields an exciton polarizability ($P$) of $(1.3\pm 0.07)\times$ $10^{-19}$ eV$\cdot$ (m/V)$^2$, consistent with previously reported values\cite{chernenko2024stark}.  
The estimated $\alpha$ is consistent with previously reported values \cite{chernenko2024stark}, however, lower than the theoretically calculated value of 2$\times$ $10^{-20}$ eV$\cdot$ (m/V)$^2$~\cite{pedersen2016exciton}. The variation of the normalized oscillator strength ($\Delta f$= $\frac{f_{V}-f_{0}}{f_{0}}$) for bare-MoSe$_2$, extracted from the fitting, is shown in Fig. \ref{DR_field}c, where  $f_0$ corresponds to zero applied voltage.  Based on the  $\alpha$ value from Fig. \ref{fig:RT_Stark shift}, the variation of absorption ($A$) is calculated with respect to electric field \textit{F}.  This calculation follows \cite{sebastian1981charge} 
\begin{equation}\label{eq:eqn-1}
	\Delta A (F) \propto  \frac{d A}{dE}  \left[ -\Delta\vec{\mu}\cdot\vec{F}  -\frac{1}{2}  \alpha F^2 \right]
\end{equation}
   
%For ML TMDC, $\frac{d \alpha}{dE} $ is assumed from the $\frac{d (\Delta R)}{dE}$. The calculated $\alpha_abs$ using Eq. \ref*{eq:eqn-1}, is shown by solid line in  \ref{gate-mapping} (i) and matches the experimental data remarkably well.
For bare-MoSe$_2$, $\frac{d A}{dE}$ at 1.57 eV is assumed to be proportional to $\frac{d (\Delta R)}{dE}$ at the same energy. The resulting $\Delta A$, calculated using Eq.\ref{eq:eqn-1} with $\Delta \mu = 0$, is shown by the solid line in Fig. \ref{DR_field}c.  It closely matches the experimental data and highlights key trends. Therefore, this change in the oscillator strength explains the variation of PL at the `off' position. A slight positive  voltage is required to achieve the maximum PL intensity by countering the intrinsic $n$-type doping caused by selenium (Se) vacancies in MoSe$_2$. 

To further corroborate the experimental observation  at the `off' position, we performed a model calculation using the transfer matrix method. The differential reflectance was obtained by considering a stack comprising the following sequence of layers: Air/\allowbreak few-layer-graphene/\allowbreak hBN/\allowbreak MoSe$_2$/\allowbreak hBN/\allowbreak few-layer-graphene/\allowbreak SiO$_2$/\allowbreak Si. % where Gr is designated as few-layer graphene.
The resulting $\Delta R/R$ values, shown in  Fig. \ref{fig:cal_R.C.}b, exhibit a variation similar to that observed in the experiment in Fig. \ref{gate-mapping}c, thereby validating the optical modeling.  In addition, a perturbative analysis is employed to quantify the influence of the electric field on exciton properties. In particular, the exciton oscillator strength is approximately proportional to the square of the electron-hole \textit{e–h} overlap. The electric field pulls electrons and holes in opposite directions, thereby reducing their overlap. The calculation reveals that the \textit{e–h} overlap decreases with increasing field strength by a factor of  $\frac{1}{ 1 + (F/F_0)^2}$, where the  characteristic field is $F_0$ = 2450 kV/cm, as shown in Fig. \ref{fig:cal_R.C.}c. This reduction in \textit{e–h} overlap directly contributes to a suppression of exciton recombination, and hence, the modulation of PL intensity observed at the `off' position in  Fig. \ref{LT-PL}c.

%Fig. \ref{gate-mapping}g suggests that in contrast to `off' position, the $\Delta R/R$ at the `on' position shows negligible variation with $V$. The near-constant $\Delta f$ at the `on' position, Fig. \ref{gate-mapping}g, implies that changes in PL intensity is not attributed to absorption variations, unlike in bare-MoSe$_2$.  This near-constant $\Delta f$ indicates that the changes in PL intensity at the `on'  position cannot be explained by absorption variations, unlike in bare MoSe$_2$. Since the reflectance data alone does not clarify the PL behavior here, we performed density functional theory (DFT) calculations to investigate the underlying electronic structure. 

\begin{figure}[t]
	\centering
	\hspace{-0.5 cm}
	\includegraphics[width=1.0\textwidth]{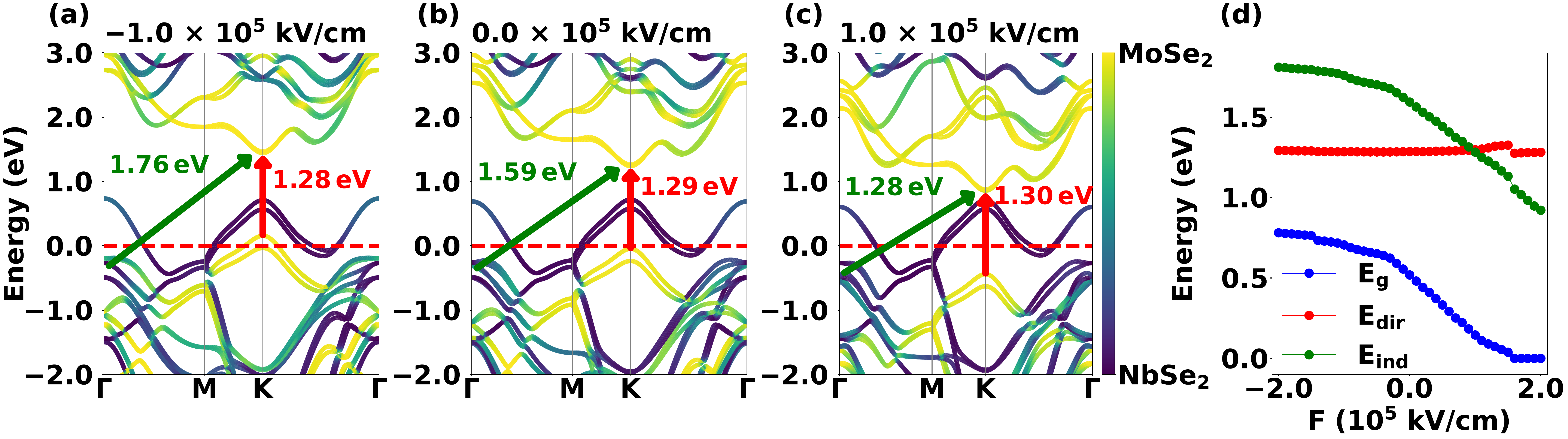}	
	%\vspace{-4.5 cm}
	\caption{ Calculated bandstructure of MoSe$_2$/NbSe$_2$ heterostructure under vertical electric fields and the corresponding band gaps: (a) F = -1.0 $\times 10^5$ kV/cm, E$_\text{g} = 0.71$ eV, (b) F = 0.0 $\times 10^5$ kV/cm, E$_\text{g} = 0.52$ eV and (c) F$ = +1.0 \times 10^5$ kV/cm, E$_\text{g} = 0.15$ eV. (d) The estimated band gap energy as an approximate measure of the optical transition energy as a function of the applied electric field. 
    %Calculated bandstructure of MoSe$_2$/NbSe$_2$ heterostructure under vertical field of (a) $F=-1$ V/\text{\AA}, (b) $F=0$ V/\text{\AA} and (c) $F=+1$ V/\text{\AA}. (d) Estimated optical transition energy as a function of applied electric field. %Schematic diagram of the MoSe$_2$/NbSe$_2$ heterostructure before contact (e), after contact with (f) $V < 0=V_0$  and, (g) V > 0 V. : At zero field ($V$ = 0 V), after optical excitation, some excitons dissociate , with free electrons transferring to NbSe$_2$. This reduces the exciton population lowering PL intensity. At $V$ > 0, the \textit{F} causes band bending, bringing excitons in MoSe$_2$ closer to electrons in NbSe$_2$. The strong spatial overlap, is supported by the small interlayer distance (~0.5 nm) relative to the single layer thickness ($\sim$ 0.7 nm). Conversely, $V$ < 0, the \textit{F} bends the bands in a way that spatially separates the holes in MoSe$_2$ from electrons in NbSe$_2$. 
		}\label{Fig-origin}
\end{figure}

Fig. \ref{DR_field}c shows that, unlike the `off' position, the $\Delta R/R$ signal at the `on' position exhibits negligible variation with $V$, indicating that changes in PL intensity are not due to absorption variations as seen in bare-MoSe$_2$. Therefore, to uncover the origin of the PL behavior at the ‘on’ position, we performed density functional theory (DFT) calculations, investigating the electronic  band structures and analyzing their impact on the intensity of bandgap transitions. 
%In contrast, the $\Delta R/R$ at the `on' position shows negligible variation with $V$, as shown  in Fig. \ref{gate-mapping}g. The near-constant $\Delta f$ at the `on' position, Fig. \ref{gate-mapping}g, implies that changes in PL intensity is not attributed to absorption variations, unlike in bare-MoSe$_2$.
%Furthermore, a perturbation calculation shows that \textit{e-h} overlap is reduced by a factor of $\frac{1}{\left( 1 + \frac{F^2}{F_0^2} \right)^2}$, where the  characteristic field is $F_0$ = 245 MV/m, as shown in Fig. \ref{fig:cal_R.C.}c. Such variation with filed justify the modification of PL intensity at `off'  position, as observed in \ref{LT-PL}c.
%Therefore, to understand  the intricate optical response of MoSe$_2$/NbSe$_2$ heterostructures in electric fields ( i.e., voltage in the experimental context), we performed density functional theory (DFT) calculations of the band structures and analyzed their impact on the intensity of bandgap transitions. 
An intense optical transition requires a direct band gap with significant contributions from MoSe$_2$ states in both valence and conduction bands. 
%An intense band gap transition requires a direct band gap with significant contributions from MoSe$_2$ states in both valence and conduction bands. 
Conversely, an indirect band gap is associated with weak band-gap emission. Hence, to address this question, bands in the vicinity of the band gap must be projected onto the parent MoSe$_2$ and NbSe$_2$ layers to ascertain the direct/indirect nature of the transition.

Our DFT-based calculations of MoSe$_2$/NbSe$_2$ heterostructures (assuming a single ML of each material) include spin–orbit coupling (SOC) and employ a localized basis set, enabling accurate treatment of strong perpendicular electric fields (see Methods). Specifically, standard plane-wave bases with supercells extended perpendicular to the layers are avoided, as they produce spurious field-induced states at the supercell boundaries.  We assume AB stacking corresponding to a single MoSe$_2$/NbSe$_2$ structural unit in the unit cell. This necessarily implies that individual layers are strained relative to their free-standing structures. Importantly, such commensurate geometries mean that MoSe$_2$ and NbSe$_2$ layers are strongly coupled electronically. Effectively, metal- and semiconductor-derived states become interlocked and, thereby, less susceptible to electric fields. Conversely, decoupled layers are highly susceptible to electric fields since the dipole moment $F\times d$, where $d$ is the center-to-center distance between the Se atoms of the two layers, will cause significant vertical shifts in the band structure. %\VKcomment{this para can be merged with above paragraph}%done

The calculated band structures are shown in Fig. \ref{Fig-origin}a-c.  In particular, the band structure in the absence of an electric field verifies that MoSe$_2$/NbSe$_2$ has a direct band gap between semiconductor-derived bands (Fig. \ref{Fig-origin}b). However, in the presence of electric fields, the indirect band gap shrinks in proportion to the field in the positive direction, while negative fields enlarge the indirect gap, c.f. Fig. \ref{Fig-origin}a and Fig. \ref{Fig-origin}c. This is in agreement with experimental spectra in Fig. \ref{gate-mapping}d. However, the field required for indirect and direct band gaps to cross is $\sim$ $1.0 \times 10^5$ kV/cm, as shown in Fig. \ref{Fig-origin}d, far greater than the experimentally observed range. We suggest that the discrepancy arises from the fact that the experimental MoSe$_2$/NbSe$_2$ heterojunctions are not commensurate but, rather, rotated relative to each other. Thereby, the electronic coupling is significantly weakened, leading to a reduction of the field required for a transition to an indirect band gap.  Furthermore, the temperature variation in the maximum PL enhancement factor at the `on' position can be  attributed to radiative recombination competing with ultrafast charge and energy transfer processes in the TMDC/metal heterostructure. At low temperatures, the reduced exciton lifetime down to a few picoseconds means that radiative recombination dominates. In contrast, at $T =$ 295 K, non-radiative channels contribute significantly, leading to a decreased PL enhancement.

\section{Conclusion}

%In summary, our findings reveal a distinctive mechanism for enhancing emission in TMDC-based van der Waals heterostructures. By applying a vertical electric field, we demonstrate the ability to tune the absolute PL intensity—by up to three orders of magnitude at the  `off' position and one order at the  `on' position. Such a large modulation is promising for applications in optical switches and modulators, as suggested in previous studies \cite{sun2016optical,shen2014electro}. The variation at the  `off' position linked to the modification of oscillator strength. In contrast, calculated band structures demonstrates that field-dependent modifications of the bandgap. Furthermore, our results indicates a significant revival of the emission at `on'  position compared to `off'.  Moreover,  temperature-dependent captures the competition between radiative recombination with charge/energy transfer between two constitutens of heterostructures. Unlike conventional tuning strategies that rely on chemical doping or strain, our approach provides clean, non-invasive control over exciton behavior through external electric fields. This work underscores the potential of \hl{utilizing} vertical electric fields as a precise, reversible, and disorder-free alternative for tailoring excitonic and electronic properties in TMDC-based heterostructures.

%In summary, our findings reveal a significant enhancement in emission in TMDC-based vdW heterostructures. 
In summary, our findings reveal a significant enhancement of emission in TMDC/metal vdW heterostructures, enabled by electric field control. By applying a vertical electric field, we demonstrate the ability to tune the absolute PL intensity--by up to three orders of magnitude at the `off' position and one order at the `on' position. Such large modulation is promising for applications in electrically controlled switches and modulators, as suggested by previous studies \cite{sun2016optical,shen2014electro}. The variation at the `off' position is linked to the modification of the excitonic oscillator strength, while calculated band structures demonstrate field-dependent modifications of the bandgap at the `on' position. %Furthermore, our results indicate a significant revival of emission at the `on' position compared to `off'. 
Temperature-dependent measurements reveal the impact of radiative recombination in PL enhancement factor. %Temperature-dependent measurements reveal a competition between radiative recombination and \hl {can we write here either charge or energy ?} charge/energy  transfer between the heterostructure’s constituents and therefore a modification in enhancement factor.
This work underscores the potential of utilizing vertical electric fields as a precise and effective method to tailor excitonic and electronic properties in TMDC-based heterostructures.  Unlike conventional tuning methods based on chemical doping or strain, our electric-field approach offers clean, reversible, and disorder-free control of exciton behavior. %Moreover, our findings \hl{can guide} open a direct route for the investigation of 2D superconductor–TMDC heterostructures below the superconducting critical temperature, potentially revealing emergent quantum states that influence both superconducting pairing and excitonic behavior.
Moreover, our findings act as a guiding light for further exploration of 2D superconductors based TMDCs heterostructures below the superconducting critical transition temperatures. This can open interesting avenues to investigate the emergent quantum states influencing superconducting pairing mechanism and the excitonic physics.
%In summary, our findings reveal a distinctive mechanism for enhancing emission in TMDC-based van der Waals heterostructures. By applying a vertical electric field, we demonstrate the ability to tune PL intensity in a 2D semiconductor–2D metal heterostructure. Our results indicate a substantial \hl{revival} %\VKcomment{can we use 'revival' word instead here ?} 
%of absolute PL intensity—by up to three orders of magnitude at the  `off' position and two orders at the  `on' position under an applied field. Such a large modulation is promising for applications in optical switches and modulators, as suggested in previous studies \cite{sun2016optical,shen2014electro}. These experimental observations are supported by calculated band structures, which show field-dependent modifications of the bandgap. Furthermore, temperature-dependent measurements reveal a variation in PL enhancement, driven by the \hl{work function difference between the constituent 2D layers}. Unlike conventional tuning strategies that rely on chemical doping or strain, our approach provides clean, non-invasive control over exciton behavior through external electric fields. This work underscores the potential of \hl{utilizing} vertical electric fields as a precise, reversible, and disorder-free alternative for tailoring excitonic and electronic properties in TMDC-based heterostructures.
% \subsection{References}
% \subsection{Floats}
% \subsection{Math(s)}
%\section{Methods}
\begin{methods}
%\textbf{Methods}\\
\noindent\textbf{Device fabrication:}
The devices were fabricated via mechanical exfoliation of bulk crystals, followed by dry transfer using a polyDimethylSiloxane (PDMS) film onto the prepatterned electrodes. Details of the electron beam lithography (EBL) process for electrode patterning are provided in the Supplementary section. Monolayer MoSe$_2$ was  obtained by exfoliation of a bulk crystal grown via chemical vapor deposition, while all other bulk crystals were purchased from HQ Graphene. The thickness of the hBN flakes was measured by an atomic force microscope. \\
\textbf{Optical spectroscopy:} Photoluminescence (PL) measurements were carried out using a continuous-wave 532 nm laser, focused to a $\sim$ 1 $\mu$m spot with a Mitutoyo objective (×50, 0.65 NA ). %For photoluminescence (PL) measurements, we used a continuous-wave green laser (532 nm) and Mitutoyo objective (×50, 0.65 NA) of spot size of $\sim$ 2 $\mu$m. 
The incident pump power was controlled using adjustable optical density filters in the excitation path. %The laser was tightly focused onto the top mirror of the cavity using a Mitutoyo objective (×50, 0.65 NA). 
For reflection measurements, a white light laser (SuperK Extreme, NKT Photonics) was used.\\
\textbf{Computational details:}
Density-functional calculations were performed using the GPAW software package \cite{gpawcite}. A localized dzp-zeta basis combined with a 15$\times$15$\times$1 k-space grid was employed, and all results are based on the PBE exchange-correlation functional including d3 van der Waals corrections. Spin-orbit interactions are included, and the AB-stacking unit cell was relaxed, yielding residual forces below 0.05 eV/\text{\AA}. The application of an electric field has a minor but significant effect on the atomic positions in the unit cell. Therefore, the structure was relaxed for all applied electric fields. The indirect bandgap is defined as the difference between the $\mathrm{K}$-point energy of the first conduction band, where MoSe$_2$ contributes most strongly, and the average $\Gamma$-point energy of the six bands immediately below the Fermi level. There are a total of six bands due to spin-orbit coupling (SOC)-induced splittings, but the resulting SOC-induced splittings at the $\Gamma$-point are negligible. We use their average energy since
all six bands exhibit some degree of hybridization, and the extent of this hybridization varies with the applied field strength.

\end{methods}

\section*{Data availability}
The data that support the findings of this study are available from the corresponding author upon reasonable request.

% of pulsed width 100-200 ps) supercontinuum as a class 4 laser source. Using a 78 Mhz

 %PL spectra were measured in backscattering configuration using a Princeton spectrometer with a 150 lines/mm (PL) and 600 lines/mm (reflection) grating. 
%A 50x objective (Mitutoyo , NA=0.65).% and the laser power was consistently maintained at 15 $\mu W$.%$\approx$ 15$\micro \watt$.  

%%%%%%%%%%%%%%%%%%%%%%%%%%%%%%%%%%%%%%%%%%%%%%%%%%%%%%%%%%%%%%%%%%%%%%
%\section{Methods}
\begin{acknowledgments}
We gratefully acknowledge funding by the Deutsche Forschungsgemeinschaft (DFG, German Research Foundation) within the projects Ho5194/16-1 and INST 93/1025-1 FUGG. S.H and A.P. acknowledge the funding from the lighthouse project IQ-Sense of the Bavarian State Ministry of Science and the Arts as part of the
Bavarian Quantum Initiative Munich Quantum Valley (15 02 TG 86). S.H., S.K. and C.G.M acknowledge financial support from the Würzburg-Dresden Cluster of Excellence on Complexity and Topology in Quantum Matter ct.qmat (EXC 2147, DFG project ID 390858490). T.G.P. was supported by the DNRF Centre CLASSIQUE sponsored by the Danish National Research Foundation, grant nr. 187. S.A.T acknowledges primary support from DOE-SC0020653 (excitonic tests on crystals). Partial support comes from NSF CBET 2330110 (environmental stability tests). S.A.T. also acknowledges partial support from Applied Materials Inc. and Lawrence Semiconductor Labs for growth systems. We are grateful for enabling us to have used the Raman measurement facility at the Julius-Maximilians-Universität Würzburg, Experimental Physics 6. %We also acknowledge the Raman measurement facility at the Julius-Maximilians-Universität Würzburg, Experimental Physics~6.

\end{acknowledgments}

\section*{Conflict of interest}
The authors declare no competing interests.

%%%%%% Bibliography
%\bibliographystyle{apsrev4-2}
%\section*{References}
\bibliography{MoSe2_PRB.bib}

\newpage
%\begin{suppinfo}
\newpage
\renewcommand{\thefigure}{S\arabic{figure}}
\renewcommand{\thetable}{S\arabic{table}}
\setcounter{figure}{0}
\setcounter{table}{0}
\setcounter{equation}{0} % If equations are used
\setcounter{enumi}{0} % For lists, if applicable
\setcounter{enumiv}{0} % For resetting the bibliography/citation numbering
\setcounter{page}{1}

\begin{center}
	\textbf{\large{Supplementary Information for}} \\
	\textbf{Electrical Control of Excitons  in Bare-MoSe$_2$ and MoSe$_2$/NbSe$_2$ Heterostructure} \\
	%"Title of the Paper" \\
	%Author1, Author2, Author3, etc.	
	Atanu Patra,$^1$	
	Vishakha Kaushik,$^1$ Ali Sepas,$^2$ Subhamoy Sahoo,$^1$
	Mathias Federolf,$^1$ Christian G. Mayer,$^{1,3}$ Sebastian Klembt,$^{1,3}$ Monika Emmerling,$^1$ Simon Betzold,$^1$
	Seth Ariel Tongay,$^4$ Fabian Hartmann,$^1$ Thomas Garm Pedersen,$^2$ and Sven Höfling$^{1,3}$

	\textit{$^1$Julius-Maximilians-Universität Würzburg, Physikalisches Institut, Lehrstuhl für Technische Physik, Am Hubland, 97074 Würzburg, Germany}\\
	 
     \textit{$^2$Department of Materials and Production, Aalborg University, DK-9220 Aalborg Øst, Denmark}\\
     
     \textit{$^3$Physikalisches Institut and Würzburg-Dresden Cluster of Excellence ct.qmat, Germany}
    
    \textit{$^4$Materials Science and Engineering, School for Engineering of Matter, Transport and
	Energy, Arizona State University, Tempe, 85287 Arizona, United States}\\
    
\end{center}
%"Title of the Paper" \\
%by Author1, Author2, Author3, etc.	

\newpage

\renewcommand{\thefigure}{S\arabic{figure}}
\renewcommand{\thetable}{S\arabic{table}}
\setcounter{figure}{0}
\setcounter{table}{0}
\setcounter{equation}{0} % If equations are used
\setcounter{enumi}{0} % For lists, if applicable
\setcounter{enumiv}{0} % For resetting the bibliography/citation numbering

\textbf{1. Supplementary Note-1}

\textbf{Pre-patterned gold-electrode} \\
%Electron beam lithography (EBL) is utilized to develop electrode patterns in silicon (Si) with high precision. The process begins with spin-coating a 1 µm thick PMMA 950K resist layer, followed by a soft bake at 165$^\circ\text{C}$ for 2 minutes to ensure proper adhesion and solvent removal. EBL is performed using a 2 nA beam current at 100 kV acceleration voltage with a dose of 1000 $\mu C/cm^2$ to define the small internal contacts. The exposed resist is developed in methyl isobutyl ketone/isopropyl alcohol (MIBK/IPA) for 45 seconds, followed by an IPA rinse to stop the development process. After patterning, SiO$_2$ etching is conducted using the Sentech RIE SI 591 system. The reactive ion etching (RIE) process employs CHF$_3$ (15 sccm) and Argon (7.5 sccm) at an RF power of 50 W for 300 seconds, achieving an etching depth of 50 nm to transfer the pattern into the SiO$_2$ layer.
\noindent Electron beam lithography (EBL) is performed to define the small internal contacts with high precision. The process begins with spin-coating a 1 µm thick PMMA 950K resist layer, followed by a soft bake at 165$^\circ$C for 2 minutes to ensure proper adhesion and solvent removal. EBL exposure is carried out using a 2 nA beam current at 100 kV acceleration voltage with a dose of 1000 $\mu C/cm^2$. After exposure, the resist is developed in MIBK/isopropanol(IPA) for 45 seconds, followed by a rinse in IPA to stop the development process. 
The patterned resist is then used for SiO$_2$ etching in the Sentech RIE SI 591 system, employing a reactive ion etching (RIE) process with CHF$_3$ (15 sccm) and Argon (7.5 sccm) at an RF power of 50 W for 300 seconds, achieving an etching depth of 50 nm. After etching, a thin metal layer is deposited, consisting of 3 nm chromium and 50 nm gold, followed by a lift-off process in methylpyrrolidone to define the contacts. 

Next, an optical positive varnish layer of 2 µm ma-P1215 is applied, followed by optical exposure and development in MA-D331. To improve adhesion and remove organic residues, an O$_2$ plasma treatment for 2 minutes is performed. For the thick outer contacts, 30 nm chromium and 300 nm gold are deposited, followed by a lift-off process in methylpyrrolidone to complete the fabrication.

%\textbf{1.Supplementary Note-2}

%Faddeeva model

\begin{figure}[h]
	\centering
	\includegraphics[width=0.75\textwidth]{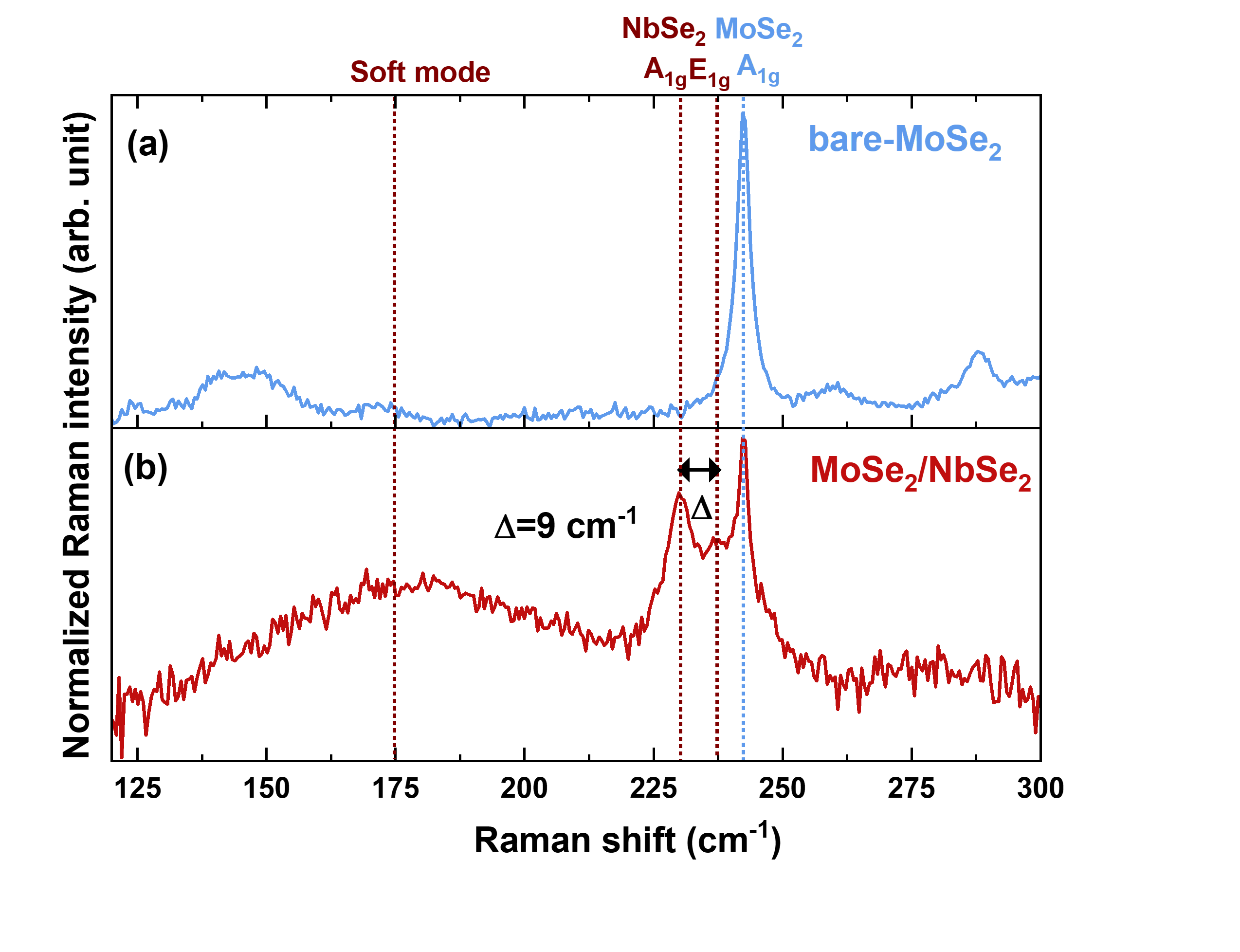}
    \vspace{-1.0 cm}
	\caption{Raman spectra of (a) bare-MoSe$_2$ and (b) MoSe$_2$/NbSe$_2$ heterostructure.}
	\label{fig:Raman}
\end{figure}

%\newpage
\textbf{Thickness characterization}\\
Figure \ref{fig:Raman}a and b show the Raman spectra of bare-MoSe$_2$ and MoSe$_2$/NbSe$_2$ heterostructures, respectively. The y-axis is presented on a logarithmic scale to better resolve the peaks. In the spectrum of bare-MoSe$_2$ (Fig. \ref{fig:Raman}a), a single peak is observed at $\sim$ 247 cm$^{-1}$, corresponding to the $A_{1g}$ mode, which is the characteristic out-of-plane vibrational mode of ML MoSe$_2$. In contrast, the spectrum of MoSe$_2$/NbSe$_2$ (Fig. \ref{fig:Raman}b) exhibits two distinct additional peaks of NbSe$_2$ namely, the $A_{1g}$ mode at $\sim$  227 cm$^{-1}$ and the $E_{1g}$ mode at 236 cm$^{-1}$. The separation between these two peaks ($\Delta$ $\sim$ 9 cm$^{-1}$)$^1$, together with the presence of a soft mode at $\sim$ 175 cm$^{-1}$, indicates that the NbSe$_2$ layer is of bulk thickness.

\begin{figure}[h]
	\centering
	\includegraphics[width=0.85\textwidth]{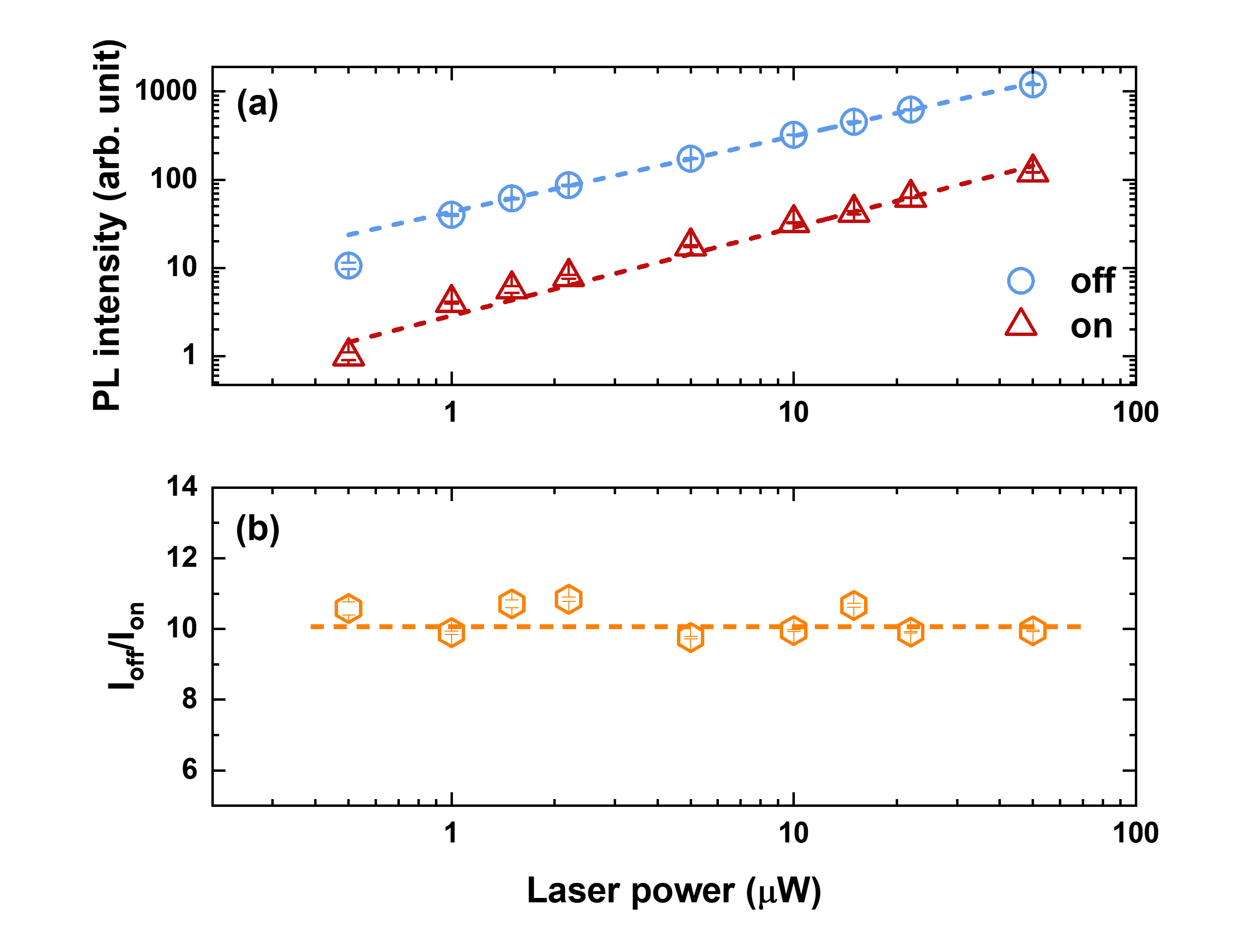}
    \caption{(a) Power-dependent PL intensity at `off' (blue) and `on' (red) positions. (b) PL quenching at the `on' position as a function of laser power, compared to `off' position.}
	%\caption{(a) Variation of PL intensity with laser power for bare-MoSe$_2$ (blue) and MoSe$_2$/NbSe$_2$ (red) heterostructure. (b) The PL quenching in heterostructure against laser power. }
	\label{fig:power dependence}
\end{figure}

\textbf{Linear regime of exciton}\\
The linear increase of PL intensity in Fig. \ref{fig:power dependence}a, both  bare-MoSe$_2$ and MoSe$_2$/NbSe$_2$ indicates that there is no loss due to exciton-exciton annihilation. 
In addition, Fig. \ref{fig:power dependence}b illustrates the PL quenching factor remains within the experimental error.

\newpage
\textbf{Voltage dependence of Device-2} 
\begin{figure}[h]
	\centering
	\includegraphics[width=0.99\textwidth]{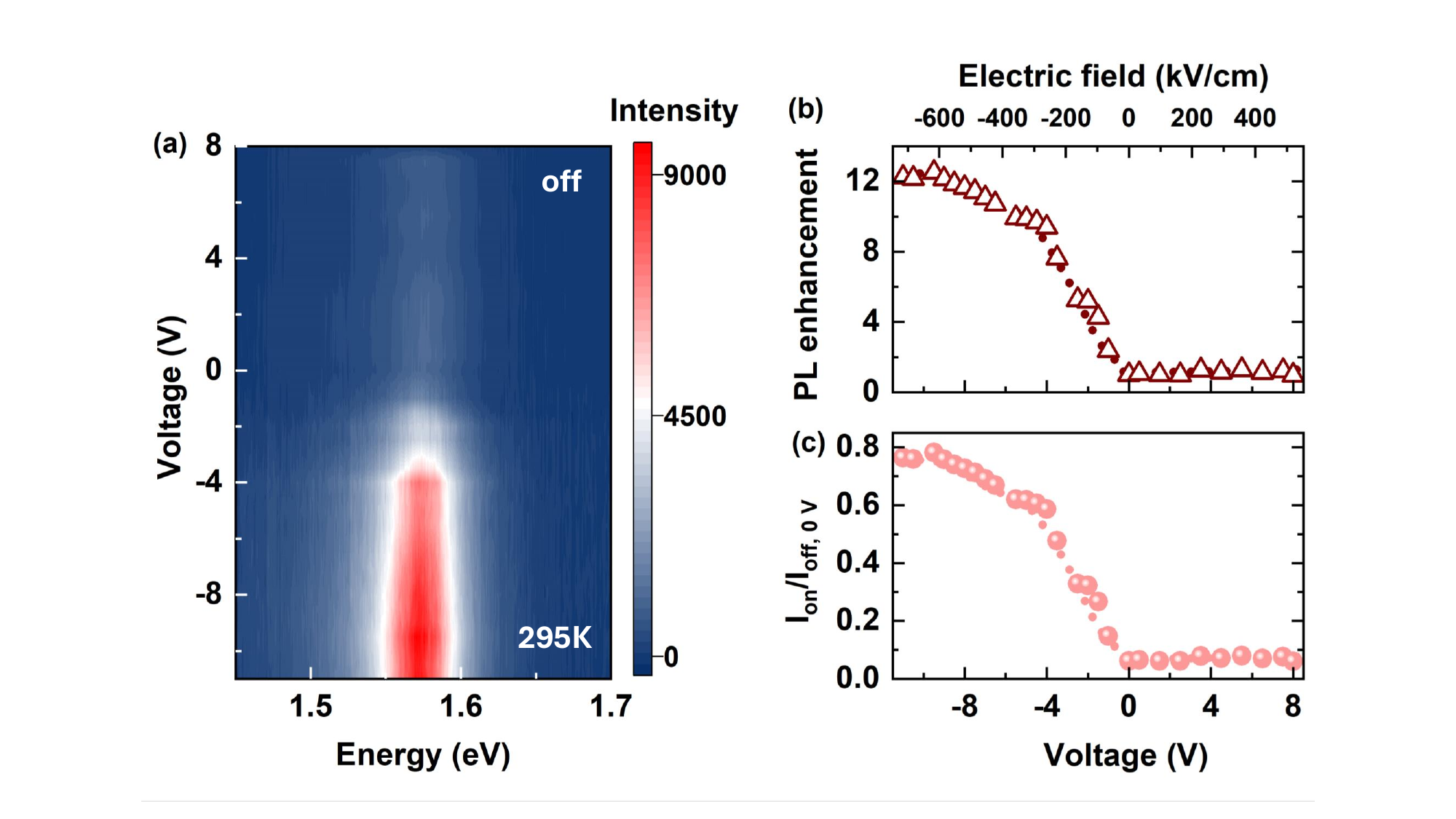}
    %\vspace*{-4.5cm}
	\caption{Voltage dependence of Device-2. (a) Contour plot of PL spectra as a function of voltage at the `on' position. (b) Enhancement of the integrated PL intensity relative to its value at $V$ = 0 V. (c) Intensity variation relative to the PL intensity at the `off' position at 0 V.}
\label{fig:V$_g$_Device_NbSe$_2$}
\end{figure}

\noindent A voltage-dependent PL enhancement is observed in device D2 of the MoSe$_2$/NbSe$_2$ heterostructure, as shown in Fig. \ref{fig:V$_g$_Device_NbSe$_2$}, similar to the behavior presented in the main text (Fig. \ref{gate-mapping}b). Figure \ref{fig:V$_g$_Device_NbSe$_2$}a shows the contour plot, and Fig. \ref{fig:V$_g$_Device_NbSe$_2$}b displays the extracted integrated PL intensity as a function of voltage. The maximum enhancement factor reaches $\sim$~12, and the PL revival is $\sim$ 80$\%$ compared to bare-MoSe$_2$, as shown in Fig. \ref{fig:V$_g$_Device_NbSe$_2$}c.

%\ref*{eqn-1}, following from the Stark effect and using the %estimated polarizabilty.

%\begin{equation}
%	\begin{aligned}
%		\Delta\alpha (E) & \stackrel{\sim}{=} \frac{d\alpha}{dE}\Delta E + \frac{1}{2}\frac{d^2\alpha}{dE^2}(\Delta E)^2 \\
%		& \stackrel{\sim}{=} \frac{d\alpha}{dE}[-\Delta\mu\cdot\vec{F} -\frac{1}{2}\Delta P F^2] + \frac{1}{2}\frac{d^2\alpha}{dE^2}(\Delta E)^2[-\Delta\mu\cdot\vec{F} -\frac{1}{2}\Delta P F^2]
%	\end{aligned}
%	\label{eqn-1}
%\end{equation}

%\newpage
\textbf{Field calculation}\\
The electric field is calculated in bare-MoSe$_2$ as$^2$ 

\[F = \frac{V}{d_{Mo} + d_h \left( \frac{\varepsilon_{\perp,0,Mo}}{\varepsilon_{\perp,0,h}} \right)} \]

%\begin{equation}
%	F = \frac{V_B}{d_{Mo} + d_h \left( %\frac{\varepsilon_{\perp,0,w}}{\varepsilon_{\perp,0,h}} \right)}
%\end{equation}

%where d$_{Mo}$$_{(h)}$ is the thickness of the ML- MoSe$_2$ (hBN), $\epsilon_{\perp,0,Mo}$  and $\epsilon_{\perp,0,h}$ are the static out-of-plane dielectric constants of MoSe$_2$ and hBN, respectively.\\
where d$_{Mo}$$_{(h)}$ is the thickness of the ML- MoSe$_2$ (hBN), $\epsilon_{\perp,0,Mo}$ (=7.2) and $\epsilon_{\perp,0,h}$(=3.76) are the static out-of-plane dielectric constants of MoSe$_2$ and hBN, respectively.\\
%\cite{laturia2018dielectric}

\begin{figure}[t]
	\centering
	\hspace*{-2.0cm}
    \includegraphics[width=1.2\textwidth]{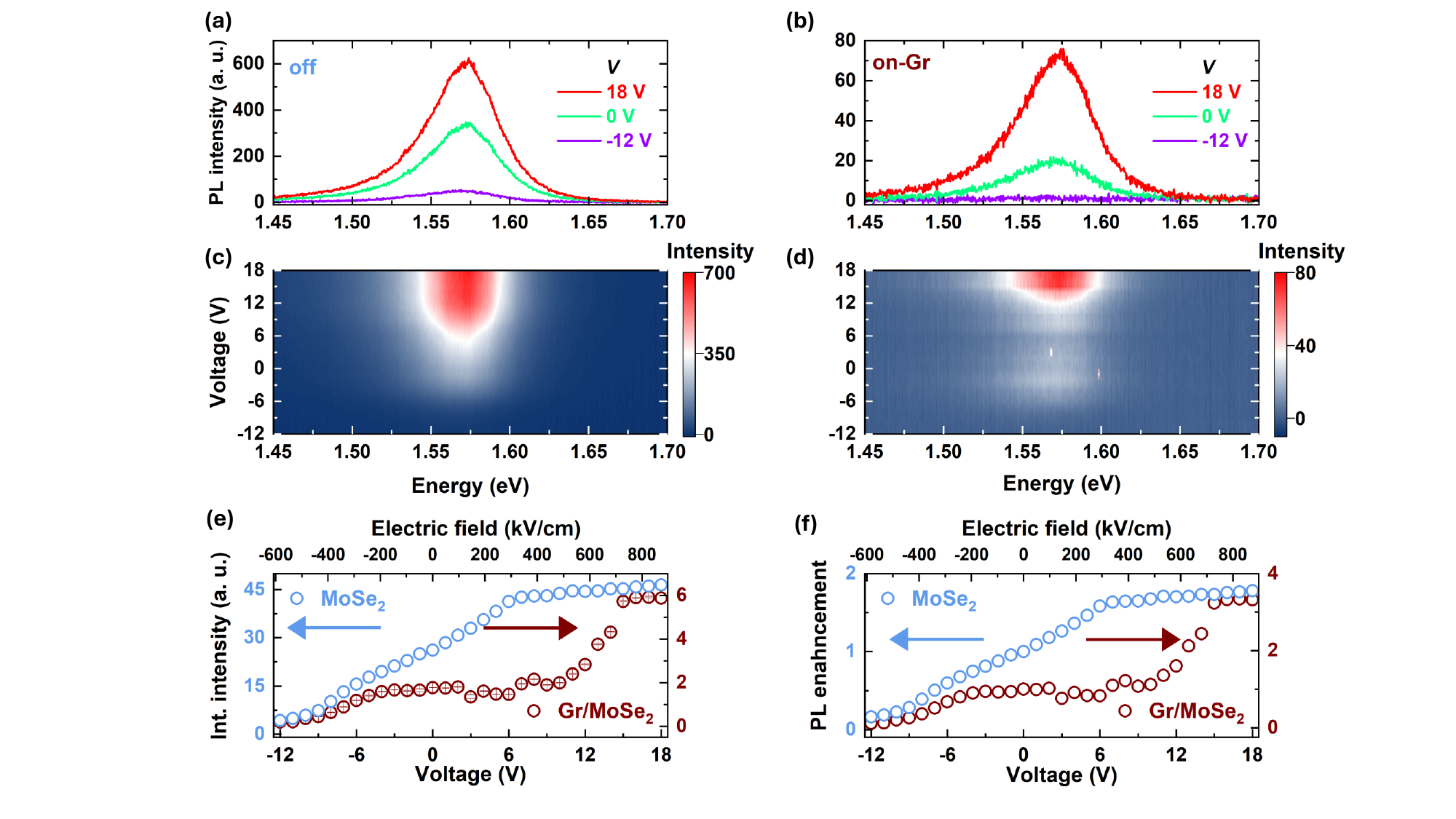}
    \vspace*{-0.5cm}
    
	\caption{Voltage dependence of PL in bare-MoSe$_2$ and FLG/MoSe$_2$ devices. (a) and (d) PL spectra at different gate voltages for the `off' and `on-graphene'  positions. (b) and (e) Corresponding contour plots as a function of voltage. (c) Integrated PL intensity for `off' (blue circles) and `on-graphene'  (red triangles) positions. (f) PL enhancement factor for the two positions.}
	\label{fig:V$_g$_Gr/MoSe$_2$}
\end{figure}

\newpage
\textbf{Voltage dependence of Device-3}\\
Another device (D3) with few-layer graphene (FLG) is placed on top of MoSe$_2$ as a metal contact instead of NbSe$_2$. Changing the position of the metal modifies the PL variation, as shown in Fig. \ref{fig:V$_g$_Gr/MoSe$_2$}. Representative PL spectra at different gate voltages for the `off' and `on-graphene' positions are shown in Fig. \ref{fig:V$_g$_Gr/MoSe$_2$}a and b, respectively. The corresponding contour plots as a function of voltage at the same positions are given in Fig. \ref{fig:V$_g$_Gr/MoSe$_2$}c and d. Importantly, in this case, PL enhancement occurs in the same direction for both positions. The integrated PL intensities are shown in Fig. \ref{fig:V$_g$_Gr/MoSe$_2$}e, with blue circles representing the `off' position and red triangles representing the `on-graphene'  position. Fig. \ref{fig:V$_g$_Gr/MoSe$_2$}f indicates that the PL enhancement factors are $\sim$ $2\times$ and $3\times$ for the `off' and `on-graphene'  positions, respectively.

\newpage

\textbf{Time-resolved measurements}

\begin{figure}[t]
	\centering
	\includegraphics[width=0.99\textwidth]{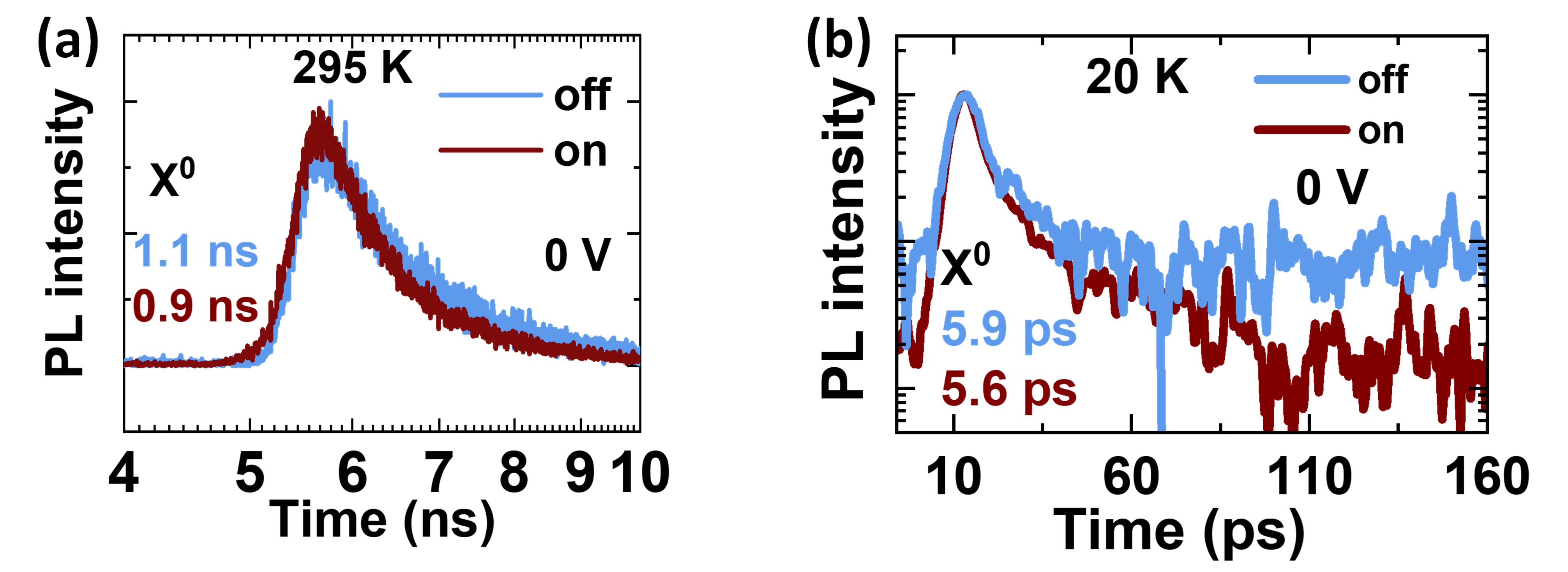}
	%\caption{Time-resolved PL measurement  at room-temperature, $T =$ 295 K (a) and 20 K (b) at `off' and `on' positions are shown in blue and red, respectively}
    \caption{Time-resolved PL measurements at room temperature ($T = 295$~K) (a) and at 20~K (b). Data at the `off' and `on' points are shown in blue and red, respectively.}
	\label{fig:Time-resolved-PL}
\end{figure}

%To perform time-resolved measurements, a different experimental setup is employed. In this setup, the sample is excited using a pulsed laser, allowing for the precise probing of the emission lifetime. 
\noindent To perform time-resolved measurements, a pulsed excitation laser is employed for precise probing of the emission lifetime. The excitation source is a Spectra-Physics Tsunami titanium-sapphire laser, which produces pulses with a duration of around 2 ps and operates at a repetition rate of 80 MHz. Since the lifetime of excitons undergoes significant changes when the sample is cooled to cryogenic temperatures, two distinct time-resolving techniques are implemented, each optimized for a different temporal range. To measure emission lifetimes within the nanosecond range, a PicoQuant $\tau$-SPAD avalanche photodiode is utilized with a time-correlated single photon counting module, PicoHarp 300. For lifetimes down to just a few picoseconds, a Hamamatsu streak camera C10910 is used. The streak camera is equipped with a charge-coupled device (CCD) detector ORCA-Lightning, which enables simultaneous spectral and temporal resolution of the emission signal. In both the techniques, precise synchronization between the recorded signal and the pulsed excitation source is required. The synchronization for the PicoHarp 300 module is directly triggered by the electronic control system of the pulsed laser, whereas the streak camera uses an optical reference signal. This is achieved by placing an optical pick-up diode along the excitation beam path, which serves as a time reference for the streak camera. To ensure high spectral resolution, both time-resolving devices are positioned after an Andor Shamrock 750 Czerny–Turner monochromator. This monochromator spectrally disperses the emitted light along one axis before it is analyzed. Additionally, a Dove prism is incorporated into the setup to rotate the image before it is directed into either the streak camera or the avalanche photodiode, ensuring proper alignment of the spectral information in the detecting system. Time-resolved PL measurement at $T =$ 295 K and $T =$ 20 K in bare-MoSe$_2$ and and MoSe$_2$/NbSe$_2$ are shown in blue and red, respectively.

\newpage

\begin{figure}[h]
	\centering
	\includegraphics[width=0.75\textwidth]{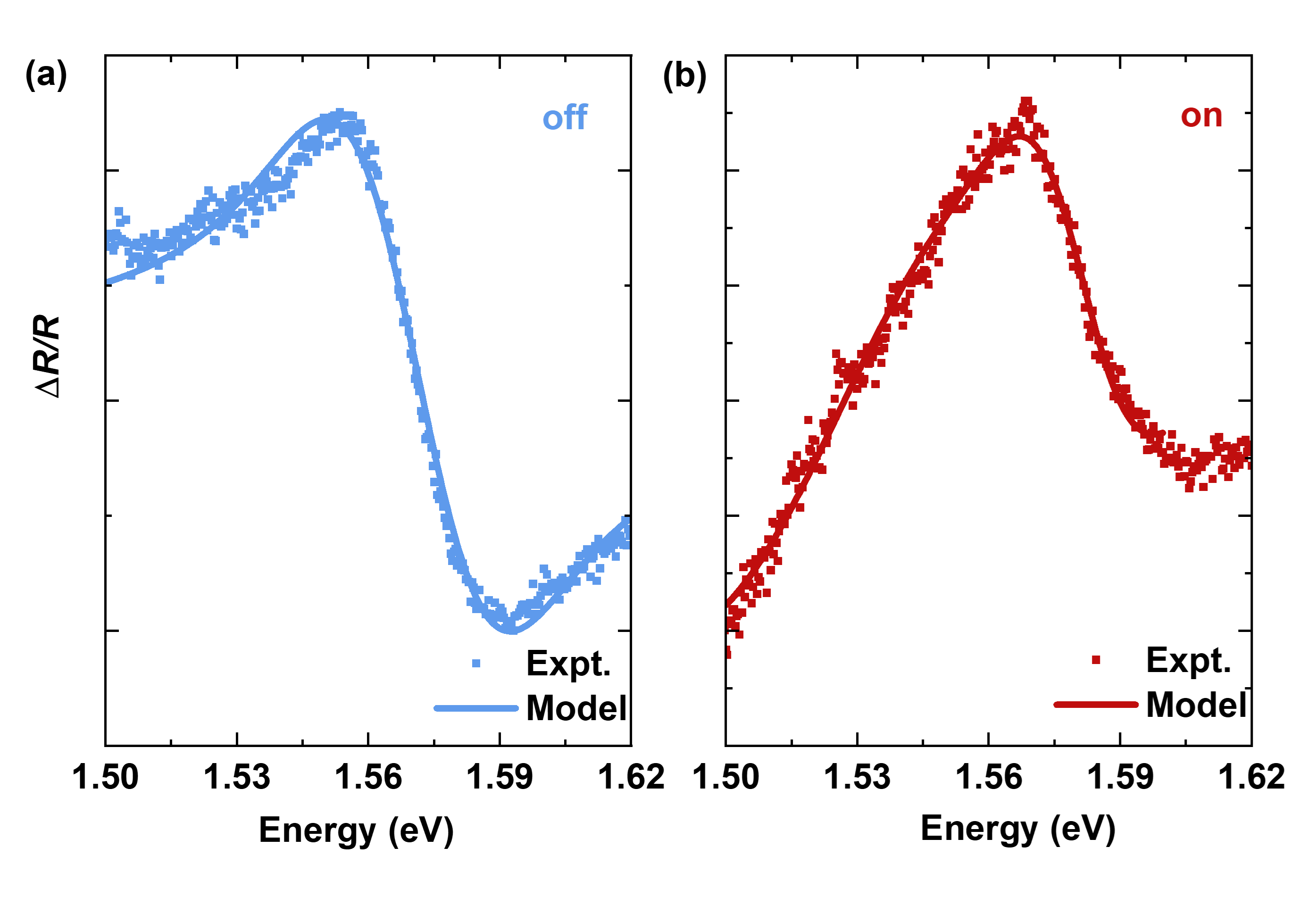}
	\caption{Fitting of the reflectance contrast ($\Delta R/R$) of bare-MoSe$_2$ (a) and MoSe$_2$/NbSe$_2$ (b) using Faddeeva model as given in ref.$^3$ }
	\label{fig:DR_fit}
\end{figure}

\textbf{Reflectance contrast analysis}\\
To quantitatively analyze the exciton resonances, we fitted the reflectance contrast ($\Delta R/R$) using the Faddeeva function, as described in ref.$^3$. Examples of the fitted spectra for bare-MoSe$_2$ and MoSe$_2$/NbSe$_2$ at $V = 0~\text{V}$ are shown in Fig. \ref{fig:DR_fit}a and b, respectively.

\newpage
\textbf{Stark shift of monolayer MoSe$_2$}

\begin{figure}[h]
	\centering
	\includegraphics[width=0.8\textwidth]{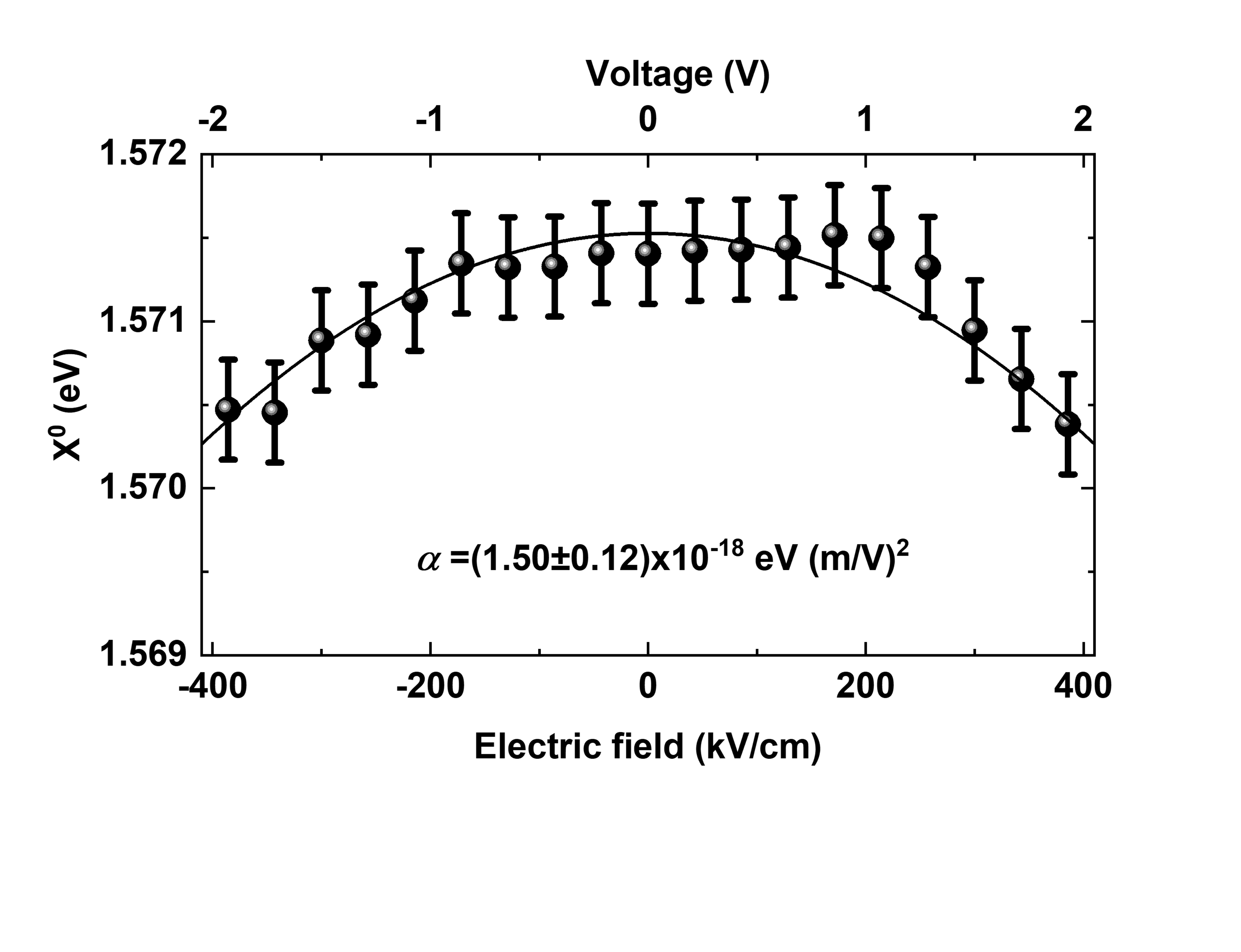}
    \vspace{-2.0 cm}
	\caption{Variation of A-exciton of bare-MoSe$_2$ with electric field. }
	\label{fig:RT_Stark shift}
\end{figure}

\noindent Variation of exciton peak position at room temperature against electric field ($F$) in bare-MoSe$_2$, as extracted from Fig. \ref{DR_field}a in the main text. A quadratic dependence is evident in Fig. \ref{fig:RT_Stark shift}, corresponding to an exciton polarizability ($\alpha$) of 1.50 $\times$ 10$^{-18}$ eV·(m/V)$^2$.\\

\begin{figure}[h]
	\centering
    %\hspace*{-3.5cm}
	\includegraphics[width=1.0\textwidth]{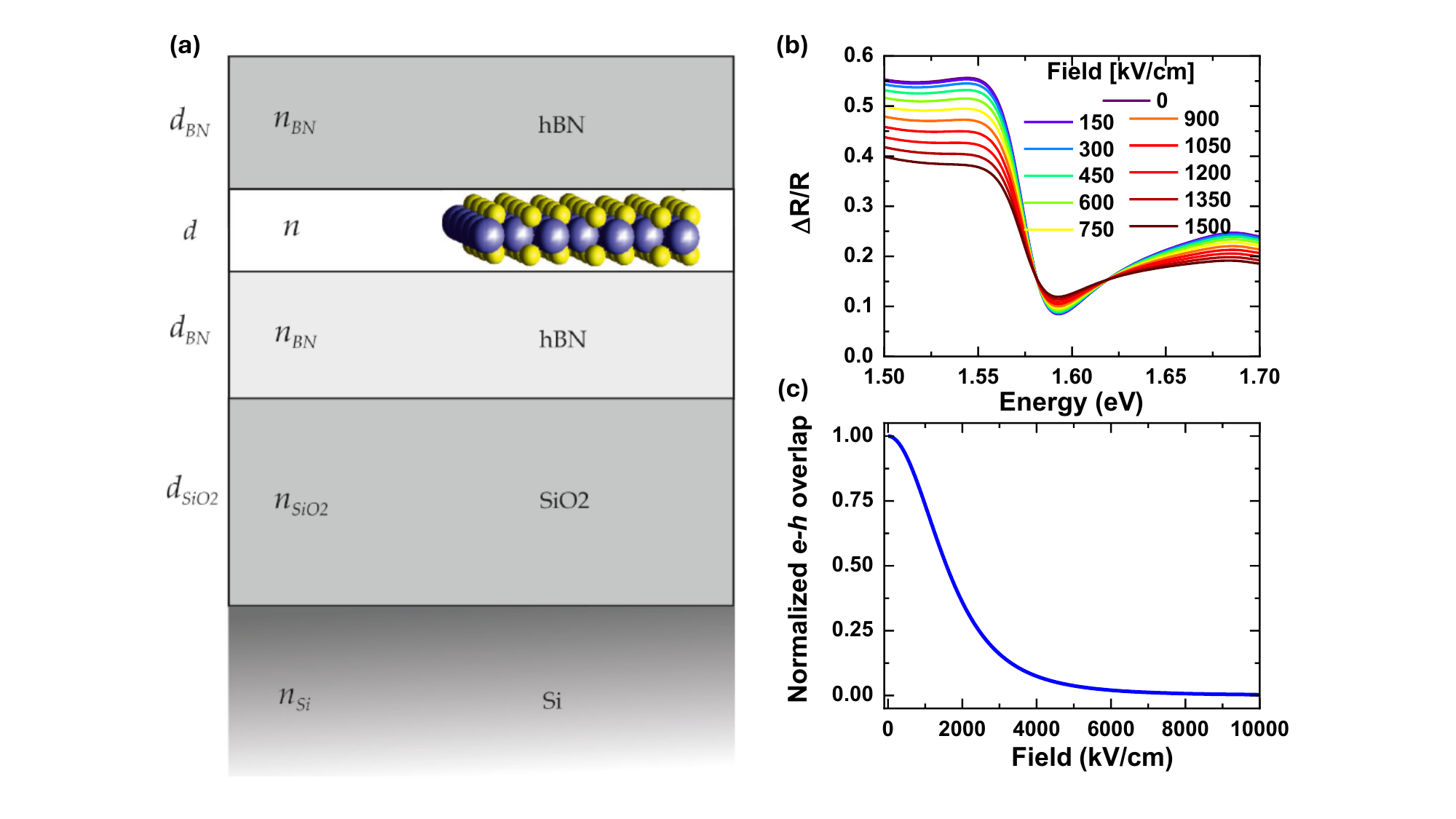}
    %\vspace*{-1.0cm}
	\caption{(a) Sketch of the model. (b) Calculated variation of reflectance contrast ($\Delta R/R$) for different electric fields. (c) Calculated \textit{e-h} overlap under vertical electric field (\textit{F}).}
	\label{fig:cal_R.C.}
\end{figure}

%\newpage
\textbf{Calculated reflectance contrast}\\
The optical properties of the stack are modelled using the transfer matrix method. We calculate the differential reflectance by considering a stack comprising the following sequence of layers: Air/Gr/hBN/MoSe$_2$/hBN/Gr/SiO2/Si, where Gr is designated as few-layer graphene. The stack is shown in Fig. \ref{fig:cal_R.C.}a.

The parameters are taken as %$d_{hBN}$=12 nm, $n_{hBN}$=2.12\\
%$d_{SiO_2}$=100 nm, $n_{SiO_2}$=1.4\\
%$n_{Si}$=3.7
\[
\begin{array}{c}
	d_{air} =\infty, \quad n_{air} = 1.0 \\
	d_{Si} =\infty, \quad n_{Si} = 3.7\\
	d_{hBN} = 12 \text{ nm}, \quad n_{hBN} = 2.12 \\
	d_{Gr} = 3.5 \text{ nm}, \quad n_{Gr} = 3+  2.5 \textit{i}\\
	d_{SiO_2} = 100 \text{ nm}, \quad n_{SiO_2} = 1.45 \\
	
\end{array}
\]

%To model the refractive index of MoSe$_2$, we use $ d = 0.63 $ nm and \( n = \varepsilon \) with  
%\[
%\varepsilon(\omega) = \varepsilon_{\infty} + \frac{g (F)}{\omega_0^2 - \omega^2 - %i\gamma\omega}
%\]

%The parameters are  
%\[
%\varepsilon_{\infty} = 6.0, \quad \omega_0 = 1.56 \text{ eV}, \quad \gamma = 0.02 \text{ eV},g (0) = 1 \text{ eV}
%\]

The MoSe$_2$ is modelled following the approach in$^4$ including excitonic and spin-orbit effects. Dielectric screening by encapsulation in hBN is incorporated using a dielectric constant of 4.5 in the Keldysh potential. In addition, a constant background contribution $\epsilon_\in$= 6 is added to the excitonic response of MoSe$_2$.

%In the presence of a perpendicular electric field \( F \), the oscillator strength becomes  

%\[
%g(F) =\frac{g(0)}{\left( 1 + \frac{F^2}{F_0^2} \right)^2} 
%\]

%A calculation using [-----] shows that the characteristic field is determined by the opposite displacements of electrons and holes in the field. For $\Delta R/R$ as shown in Fig. \ref{fig:cal_R.C.} (b), we use $F_0$ = 200 MV/m.

The modification of the dielectric properties of MoSe$_2$ by a perpendicular external electric field is accounted for by including the reduced overlap between electrons and holes. Within the exciton envelope approximation, the exciton oscillator strength is determined by the electron-hole overlap. In an electric field, the carriers are pulled in opposite directions, leading to a reduced overlap. We model the effect by considering carriers in a 0.63 nm wide quantum well with finite barriers of 0.5 eV. The effective masses are assumed identical and takes as twice the reduced mass for direct excitons in MoSe$_2$, see$^5$. A perturbation calculation shows that oscillator strengths are reduced by a factor of $\frac{1}{\left( 1 + \frac{F^2}{F_0^2} \right)^2}$, where the  characteristic field is $F_0$ = 2450 kV/cm. \\

%The leakage current measurement, performed using a Keithley source meter at the same port as the applied  voltage ($V$), is shown in Fig. \ref{fig:leakage_current}.

%\newpage
\textbf{Leakage  current}

\begin{figure}[h]
	\centering
	\includegraphics[width=0.9\textwidth]{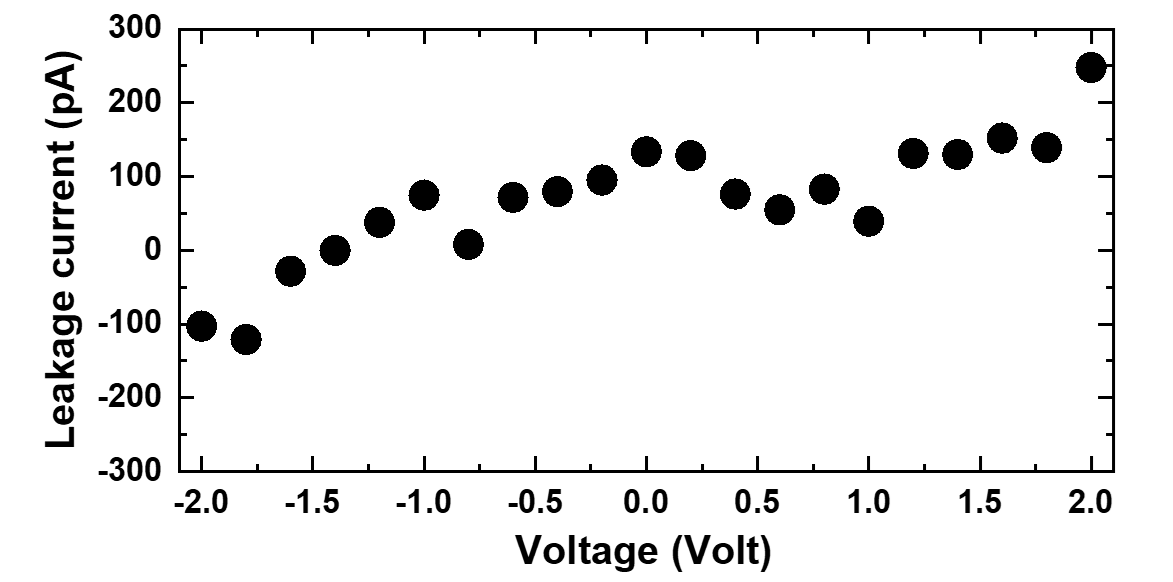}
	\caption{Leakage current measured in Device-1.}
	\label{fig:leakage_current}
\end{figure}
\noindent Another possibility of field dependence of PL intensity at the `on' positions  could be leakage current within the system. A $V>0$, i.e. electrode near to MoSe$_2$ has a high potential, and the electrode close to NbSe$_2$ is at ground potential, induces a leakage current from top to bottom, corresponding to electron flow from NbSe$_2$ to MoSe$_2$. The work function of NbSe$_2$ lies below the valence band maximum (VBM) of MoSe$_2$. Hence, a positive current enhances the electron density in MoSe$_2$, converting neutral excitons to trions and reducing PL intensity. Additional free charge carriers increase Coulomb screening, lowering exciton binding energy and destabilizing excitons, leading to further PL quenching. Excess electrons may also enhance non-radiative pathways, such as Auger or defect-assisted recombination. Conversely, for $V<0$, the current flow reverses direction, i.e., electrons flow from MoSe$_2$ to NbSe$_2$. Even for this condition, excitons are dissociating and therefore expected to exhibit PL reduction, contradicting the observed behavior. Furthermore, no significant leakage current is observed (see Fig. \ref{fig:leakage_current}).

\clearpage  % Ensures all figures appear before the reference section
\section*{References}  % No numbering for supplementary references

\begin{enumerate}

    \item H. M. Hill, A. F. Rigosi, S. Krylyuk, J. Tian, N. V. Nguyen, A. V. Davydov,D. B. Newell, and A. R. Hight Walker; Comprehensive optical characterization of atomically thin NbSe$_2$, Phys. Rev. B  \textbf{98}, 165109 (2018).

	\item S. Das, M. Dandu, G. Gupta, K. Murali, N. Abraham, S. Kallatt, K. Watanabe, T. Taniguchi, K. Majumdar,  Highly Tunable Layered Exciton in Bilayer WS$_2$: Linear Quantum Confined Stark Effect versus Electrostatic Doping, ACS Photonics \textbf{7}, 3386–3393 (2020).

    \item A. Raja, L. Waldecker, J. Zipfel, Y. Cho, S. Brem, J. D. Ziegler, M. Kulig, T. Taniguchi, K. Watanabe, E. Malic,  T. F. Heinz, T. C. Berkelbach, A. Chernikov,  Dielectric disorder in two-dimensional materials, Nat. Nanotechnol. \textbf{14}, 832–837 (2019).

	\item T. G. Pedersen,  S. Latini, K. S. Thygesen, H. Mera, B. K. Nikolić, Exciton ionization in multilayer transition-metal dichalcogenides, New J. Phys. \textbf{18}, 073043 (2016).

	\item A. Taghizadeh, T. G. Pedersen,  Nonlinear optical selection rules of excitons in monolayer transition metal dichalcogenides, Phys. Rev. B \textbf{99}, 235433 (2019).
    
\end{enumerate}

\end{document}